\def\paperversion{camera}

\def\paperversiondraftdraft

\def\paperversiondraft{draft}
\def\paperversionblind{blind}
\def\paperversioncamera{camera}
\ifx\paperversion\paperversiondraft
\else
  \ifx\paperversion\paperversionblind
  \else
    \ifx\paperversion\paperversioncamera
    \else
       \errmessage{paperversion must be set to 'draft', 'blind', or 'camera'}
    \fi
  \fi
\fi

\def\grammarlyon{on}

\documentclass[letterpaper,10pt,conference]{IEEEtran}

\usepackage{cite}
\usepackage{amsmath,amssymb,amsfonts}
\usepackage{algorithmic}
\usepackage{graphicx}
\usepackage{textcomp}
\usepackage{xcolor}
\usepackage{xargs}
\usepackage{todonotes}
\usepackage{xparse}
\usepackage{xifthen, xstring}
\usepackage{ulem}
\usepackage{xspace}
\usepackage[hidelinks=true]{hyperref}
\usepackage[T1]{fontenc}
\usepackage{colortbl}
\usepackage{comment}

\usepackage{balance}
\usepackage{multirow}

\usepackage{listings}
\makeatletter
\font\uwavefont=lasyb10 scaled 652
 
\newcommand\colorwave[1][blue]{\bgroup\markoverwith{\lower3\p@\hbox{\uwavefont\textcolor{#1}{\char58}}}\ULon}
\makeatother

\ifx\paperversion\paperversiondraft
\newcommand\createtodoauthor[2]{%
\def\tmpdefault{emptystring}
\expandafter\newcommand\csname #1\endcsname[2][\tmpdefault]{\def\tmp{##1}\ifthenelse{\equal{\tmp}{\tmpdefault}}
   {\todo[linecolor=#2!20,backgroundcolor=#2!25,bordercolor=#2,size=\scriptsize]{\textbf{#1:} {##2}}}
   {\ifthenelse{\equal{##2}{}}{\colorwave[#2]{##1}\xspace}{ \todo[linecolor=#2!10,backgroundcolor=#2!25,bordercolor=#2,size=\scriptsize]{\textbf{#1:} ##2}\colorwave[#2]{##1}}}}}
\else
\newcommand\createtodoauthor[2]{%
\expandafter\newcommand\csname #1\endcsname[2][\@nil]{}}
\fi

\ifx\paperversion\paperversiondraft
  \paperwidth=\dimexpr \paperwidth + 6cm\relax
  \oddsidemargin=\dimexpr\oddsidemargin + 3cm\relax
  \evensidemargin=\dimexpr\evensidemargin + 3cm\relax
  \marginparwidth=\dimexpr \marginparwidth + 3cm\relax
\fi

\createtodoauthor{grosser}{red}
\createtodoauthor{gysi}{blue}
\createtodoauthor{htor}{green}
\createtodoauthor{muellch}{orange}
\createtodoauthor{oli}{purple}
\createtodoauthor{az}{cyan}
\createtodoauthor{wicky}{brown}

\usepackage{booktabs}   
\usepackage{subcaption} 

\def\BibTeX{{\rm B\kern-.05em{\sc i\kern-.025em b}\kern-.08em
    T\kern-.1667em\lower.7ex\hbox{E}\kern-.125emX}}
\begin{document}
\IEEEoverridecommandlockouts

\definecolor{ourgreen}{HTML}{a8ddb5}
\definecolor{ourgray}{HTML}{b7b7b7}
\definecolor{ourblue}{HTML}{43a2ca}

\newcommand*{\affmark}[1][*]{\textsuperscript{#1}}

\author{
	\IEEEauthorblockN{Tobias Gysi\affmark[1], Christoph M\"uller\affmark[1], Oleksandr Zinenko\affmark[2], Stephan Herhut\affmark[2],
Eddie Davis\affmark[3], Tobias Wicky\affmark[3]\\
Oliver Fuhrer\affmark[3], Torsten Hoefler\affmark[1], Tobias Grosser\affmark[1]}
\IEEEauthorblockA{ETH Zurich\affmark[1], Google\affmark[2], Vulcan\affmark[3]\\
\{gysit, muellch, htor, tgrosser\}@inf.ethz.ch, \{zinenko, herhut\}@google.com, \{eddied, tobiasw, oliverf\}@vulcan.com  }
} 


\title{Domain-Specific Multi-Level IR Rewriting for GPU\\
{\Large The Open Earth Compiler for GPU-Accelerated Climate Simulation}}

\maketitle
\ifx\grammarly\grammarlyon
  \onecolumn
\else
\fi

\begin{abstract}

Traditional compilers operate on a single generic intermediate representation (IR). These IRs are usually low-level and close to machine instructions. As a result, optimizations relying on domain-specific information are either not possible or require complex analysis to recover the missing information. In contrast, multi-level rewriting instantiates a hierarchy of dialects (IRs), lowers programs level-by-level, and performs code transformations at the most suitable level. We demonstrate the effectiveness of this approach for the weather and climate domain. In particular, we develop a prototype compiler and design stencil- and GPU-specific dialects based on a set of newly introduced design principles. We find that two domain-specific optimizations (500 lines of code) realized on top of LLVM's extensible MLIR compiler infrastructure suffice to outperform state-of-the-art solutions. In essence, multi-level rewriting promises to herald the age of specialized compilers composed from domain- and target-specific dialects implemented on top of a shared infrastructure.
\end{abstract}

\section{Introduction}

\muellch{Had a hiccup reading the first sentence of the introduction. The first verb is in the present ``are revolutionizing'', the second verb is in the past ``sparked''. For me personally this did not flow well.}
Domain-specific approaches are revolutionizing the generation of high-performance
device-specific code and sparked the development of powerful domain-specific
language (DSL) frameworks, often achieving performance numbers unattainable for
general-purpose compilers~\cite{mullapudi2015polymage, tianqi2018tvm,
rawat2015SDSLc,yask2016, sourouri2017panda,tang2011pochoir}. For example,
Halide~\cite{ragan2013halide} automated the generation of high-performance code
for image processing, XLA~\cite{leary2017xla} exploited domain-specific
compilation to accelerate deep learning, and Stella~\cite{gysi2015stella} was
the first to move the weather and climate simulation to GPUs leading to
2.9$\times$ speedup~\cite{fuhrer2014stella}.

The broad success of domain-specific compilers---over time---also exposed their
largest weakness: their one-off implementations mostly separated from
general-purpose production compiler pipelines. Halide, XLA, Stella, and others
are specialized solutions for their respective domains that are not designed
with reusability in mind. The small number of reusable compiler infrastructures,
research-oriented such as ROSE~\cite{quinlan2000rose} or production such as
LLVM~\cite{lattner2004llvm}, evidences of a significant effort required to
design and maintain the infrastructure compared to implementing domain-specific
functionality. As a result, the ongoing trend of designing standalone DSL compilers compartmentalizes
the developer communities, spreads the efforts, hinders innovation transfer, and leads us
to ask: ``how can we design a domain-specific compiler that (a) is
cleanly decoupled from user-facing front-ends, (b) makes it easy to implement
domain-transformations, and (c) clearly separates potentially generic
components?''

\begin{figure}
  \centering
  \includegraphics[width=.98\columnwidth]{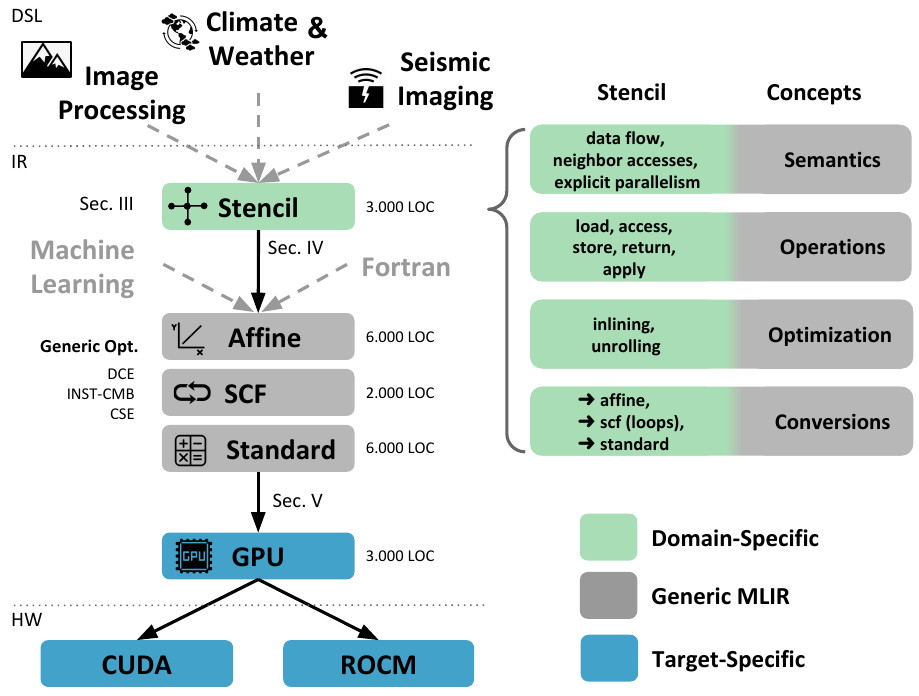}
  \caption{The Open Earth Compiler.}
  \label{fig-overview}
\end{figure}


We take a practical case study-based approach to addressing this question by designing and implementing a domain-specific compiler for weather and climate modeling. While this domain uses the stencil computational pattern found in image processing~\cite{ragan2013halide} or seismic imaging~\cite{seismic}, it often requires radically different optimization strategies to reach maximum performance~\cite{gysi2015stella}. Weather and climate models~\cite{harris2013fv3dycore,cosmo} operate on 3D domains and execute bandwidth-limited low-order stencils containing control flow. In comparison, image processing pipelines apply regular stencils to 2D data structures, and seismic imaging stencils are high-order and compute-intensive.
On the other hand, the
underlying abstraction of loops over multi-dimensional arrays, the arithmetic
optimizations, and the conversion to device-specific GPU code is mostly
identical across these domains, yet often reimplemented~\cite{sujeeth2014delite}.

We propose to design DSL compilers using \emph{multi-level IR rewriting}. This
approach is a combination of (a) intermediate representations (IR)
based on Static Single Assignment form (SSA)~\cite{rosen1988global}, (b)
operations with high-level semantics, and (c) progressive lowering, which provides an
effective framework for reusable domain-specific high-performance code generation.
SSA-based IRs allow us to reuse optimizations from general-purpose
compilers~\cite{muchnick1998compilers}.  High-level operations concisely encode
domain properties and make them readily available as, e.g., SSA data flow
without a need for costly analyses. Progressive lowering makes it natural to
preserve domain information, to express transformations as high-level peephole
optimizations~\cite{mckeeman1965peephole}, and to introduce reusable
lower-level abstractions. The recently introduced MLIR compiler
infrastructure~\cite{lattner2020mlir} allows us to instantiate
production-quality compiler IRs that follow the practice-proven IR design
principles developed in LLVM~\cite{lattner2004llvm} over the last
15 years.

The Open Earth Compiler\footnote{\url{https://github.com/spcl/open-earth-compiler/}} we implemented (\autoref{fig-overview}\grosser{confirm
logo use before publication. We can leave these logos during the review
process.}\gysi{all of mlir is 95.000 LOC!})
is the first end-to-end compilation flow that leverages multi-level
IR rewriting for high-performance code generation.
Its core consists of a set of MLIR dialects, i.e.,
collections of domain-specific operations and transformations, and conversions
between them. The Open Earth Compiler optimizes programs by progressively
converting them from higher-level domain-specific dialects to lower-level
platform-specific ones, using peephole-style rewrite patterns.
Each dialect defines an abstraction that makes relevant
analyses inexpensive and transformations convenient to implement.

The compilation process starts from the stencil dialect (\autoref{sec-dialect}) designed as a target for various user-facing DSLs as well as a data structure for domain-specific transformations such as stencil inlining (\autoref{sec-opttrafos}).
\gysi{Do we want to make the point that developers can use the
	DSL of their choice?}
\wicky{I think saying this allows for user-specific / model specific languages is valuable as it would close the arc to the intro}
Stencils are then lowered through a series of IRs featuring index operations (Affine), 
structured control flow (SCF), and arithmetic operations (Standard), 
all of which are readily available in MLIR~\cite{lattner2020mlir},
together with loop- and value-level transformations such as unrolling or common subexpression elimination.
These IRs let us target a structured loop abstraction instead of low-level
``goto''-based SSA IR commonly found in compiler backends. We use this
structure to design a generic GPU kernel dialect (\autoref{sec-gpulow})
and to implement loop-to-kernel conversion using simple patterns based on the parallelism information
preserved from the stencil level, thus avoiding expensive GPU mapping
algorithms~\cite{grosser2016pollyacc,verdoolaege2013ppcg}. The complete
pipeline transforms our high-level climate-code into a fast target-specific
binary.

While the stencil dialect is generic enough to cover a
range of applications (e.g., image processing or seismic imaging), our 
focus is excellent performance for the climate domain. The
semantics of our stencil operations enable us to replace complex
sequences of loop transformations with  generic instruction-level transformations, e.g., redundancy elimination, requiring little analysis to ensure
validity.
Other domains can adapt our stencil dialect to their transformation needs, or
reuse only the mid- and low-level abstractions. We demonstrate that thanks
to the multi-level IR rewriting, developing a domain-specific compiler
with reusable components
	is surprisingly simple provided a sufficiently expressive infrastructure.

	~\\
Our contributions are:
\begin{itemize}
  \item An approach to designing a modular domain-specific compiler using
	  multi-level IR rewriting (\autoref{sec-mlrewriting}).
  \item A stencil language expressed as an MLIR dialect, which
	encodes the high-level data flow of a stencil program as SSA def-use
  chains (\autoref{sec-dialect}).
  \item A set of transformations to tune stencil programs at a high level
	using simple peephole optimizations instead of conventional loop
	transformations (\autoref{sec-stenciltrafos}).
  \item A separate-compilation scheme for GPUs based on a platform-neutral GPU dialect
	convertible to vendor-specific GPU code (\autoref{sec-gpulow}).
  \item An evaluation on benchmarks relevant to real climate
        models: COSMO (Europe) and FV3 (US) (\autoref{sec-eval}).
\end{itemize}

\section{Multi-Level IR Rewriting}
\label{sec-mlrewriting}
Multi-level IR rewriting promises to simplify the development of domain-specific compilers by defining a stack of reusable abstractions and by implementing the transformation at the most relevant level. The goal is to minimize the complexity of each level and to reduce the cost of analysis by encoding and preserving transformation validity preconditions directly in the IR.

Identifying pertinent domain-specific abstractions is paramount to the multi-level rewriting. Each new abstraction increases risks of excessive complexity or, on the contrary, of incompleteness where some workloads cannot be represented. We instantiate
the Open Earth Compiler using the MLIR infrastructure~\cite{lattner2020mlir}, carefully considering the abstractions it provides and
introducing new ones when necessary. Our objective is to facilitate performance extraction from one of the two primary sources: parallelism and data locality, which often require conflicting transformations yielding complex optimization problems~\cite{zinenko2018scheduling}. Instead of attacking these problems frontally, we design abstractions so as to extract parallelism and locality information from the domain knowledge, using the following principles.
\muellch{``attacking these problems frontally'' - a little too colorful?}


%
%

\textbf{P1 Transformation-Driven Semantics.}
The domain abstractions for different levels of our pipeline, e.g., stencils or GPU kernels, should favor transformation-readiness over programmer-friendliness. Our objective is to build a stack of intermediate representations that enable the compiler to reason about domain-specific programs without resorting to complex analyses such as loop extraction~\cite{grosser2012polly} or dependence analysis~\cite{vasilache2006violated}. Each IR in the stack is focused on a specific set of domain transformations and designed to make all necessary information readily available. End-user usability aspects are deliberately deferred to DSL front-ends.

\textbf{P2 Progressive Lowering.}
We aim for an effective and streamlined transformation pipeline where programs are progressively lowered~\cite{lattner2020mlir} from a high-level domain IR to a low-level target IR. The different IR abstractions should be designed to maintain high-level semantic information as long as necessary, such that a potentially complex recovery of high-level concepts can be avoided. An important additional aspect of progressive lowering in larger domain-specific compilers is that abstractions should seamlessly compose with each other to coexist in a single module while the lowering is applied selectively.

\textbf{P3 Explicit Separation.}
Given the abstraction composability mandated by the previous principles, it is easier to combine individual pieces of the abstraction than to disentangle a complex representation. Our incarnation of the ubiquitous separation-of-concerns approach relies on the domain-relevant separation being explicit in at least \emph{some} intermediate abstraction in our stack. In particular, performance-related aspects of the abstractions, such as the degree of parallelism or the memory footprints, should be present in the IR and should be modifiable separately from each other. Similarly, compile- and run-time aspects of the abstraction should be separated. In the longer term, such representations are better amenable to modern search techniques~\cite{beaugnon2017optimization,adams2019halideopt,gysi2019absinthe}.

\begin{table}[t]
	\centering
	\footnotesize
	\begin{tabular}{lllr}
		\toprule
		\textbf{Level}         & \textbf{Concepts}  & \textbf{Transformations}                         & \textbf{Sec.}                   \\
		\hline
		\rowcolor{ourgreen!40}
		& - parallel stencil evaluation & & \\
		\rowcolor{ourgreen!40}
		& - value semantic  &
		&                                    \\
		\rowcolor{ourgreen!40}

		&  - explicit data flow &            &                                    \\
		\rowcolor{ourgreen!40}
		& - compile-time access offset        &                                                  &                                    \\
		\rowcolor{ourgreen!40}

\multirow{-5}{*}{Stencil} 
		& - compile-time domains        &  \multirow{-5}{*}{\shortstack[l]{- inlining (+CSE) \\ - unrolling (+CSE)}}                                                 & \multirow{-5}{*}{\shortstack[l]{ \ref{sec-dialect} \\ \ref{sec-opttrafos} \\ \ref{sec-preptrafos} }} \\
		\hline

		\rowcolor{ourgray!40}
		Standard \&                & - multi-dimensional storage     &                                    & \\
		\rowcolor{ourgray!40}
		Affine \&              & - affine index computation  &                                     &                                    \\
		\rowcolor{ourgray!40}
		SCF                 & - parallel loop  &   \multirow{-3}{*}{\shortstack[l]{- loop mapping \\ - loop to GPU}}                                               &      \multirow{-3}{*}{\ref{sec-stenlow}}                               \\

		\hline

		\rowcolor{ourblue!40}
		    & - host/device code  & - GPU outlining                                  &  \\

		\rowcolor{ourblue!40}
\multirow{-2}{*}{GPU} 
		& - SIMT parallelism  & - host/device comp.                              &      \multirow{-2}{*}{\ref{sec-gpulow}}                               \\
		\hline
	\end{tabular}
	\caption{Domain-specific to device-specific abstractions.}
	\label{tab-abstractions}
\end{table}

The abstractions we use enable progressive lowering (P2) from domain-specific to device-specific concepts providing clear separation (P3) between levels. As listed in \autoref{tab-abstractions}, each level makes specific transformations easy to implement (P1). This multi-level representation also helps us separate optimizing transformations from the lowering between the levels that constitute a large portion of DSL compilers.

\section{The MLIR Infrastructure}
\label{sec-mlir}
MLIR is a recent production compiler infrastructure that is particularly well-suited for multi-level IR rewriting thanks to its extensibility through \emph{dialects} and its built-in support for declarative rewrite patterns~\cite{lattner2020mlir}. The Open Earth Compiler can be thus implemented as a set of MLIR dialects, and transformations as rewrite patterns. Furthermore, we can readily reuse the Standard, Structured Control Flow, Affine, and LLVM IR dialects if we design our abstractions so that they compose with these.
\muellch{``thanks to'' - maybe find a more formal expression}

Core MLIR concepts include operations, values, types, attributes, (basic) blocks, and regions. An \emph{operation} is an atomic unit of program description. A \emph{value} represents data at runtime and is always associated with a type known at compile time. Operations use values (but do not consume them) and define new values. Values can only be defined once, making the IR obey SSA form. A \emph{type} holds compile-time information about a value, while \emph{attributes} provide a way to attach compile-time information to operations. A \emph{block} is a sequence of operations that, together with other blocks, connects to regions. A \emph{region} is attached to an operation that defines its semantics. Non-trivial control flow is only allowed between an operation and the regions attached to it, and between the entry and exit points of blocks that belong to the same region. Specific operations define the structure of the control flow, for example, the last operation in a block (a terminator) can conditionally or unconditionally transfer the control flow to another block.

\begin{figure}
	\centering
  \includegraphics[width=\columnwidth]{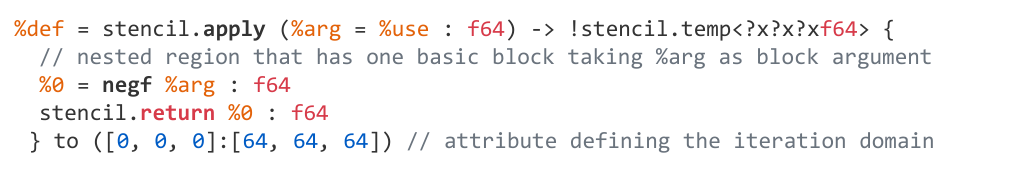}
	\caption{Example MLIR operation that sets 64x64x64 elements of the defined value \textbf{\%def} to the negative of the value \textbf{\%use}.}
	\label{fig-mlir}
\end{figure}

\autoref{fig-mlir} illustrates the syntax for an example operation from our Stencil dialect. The stencil.\textbf{apply} operation uses a value \textbf{\%use} and defines a value \textbf{\%def}. Types and attributes annotate the operation with compile-time information such as the iteration domain. The nested region consist of a single basic block that implements computation performed by the operation using the basic block argument \textbf{\%arg}. This hierarchical organization into blocks and regions enables infinite nesting.

There is no fixed set of operations, attributes, or types. Instead, each MLIR user can define their own or reuse those defined by others. Even MLIR's built-in  functionality heavily relies on this extensibility. For example, a function is a regular operation with a region containing the function body instead of a concept on its own. Thus stencil computations can be made first-class in a stencil compiler by providing custom types, attributes, and operations. The same holds for control flow operations such as loops or data types such as multi-dimensional arrays. From an infrastructure point of view, custom operations and types are indistinguishable from common scalar operations and types.

A dialect is a set operations, attributes, and types designed to work together. There is no formal or technical restriction on how dialects are structured.
Unless prescribed otherwise by the semantics of the operation, a region can contain operations from different dialects, and an operation can reference types and attributes defined by a different dialect. Therefore, new abstractions can be introduced into the MLIR ecosystem as new dialects.


\section{The Stencil Dialect}
\label{sec-dialect}


The Open Earth Compiler operates on weather and climate models. These models integrate partial differential equations forward in time commonly using either finite difference or finite volume discretization. They comprise dozens of \emph{stencil programs} (\autoref{sec-stencil-programs}) consisting of multiple dependent \emph{stencil operators} (\autoref{sec-stencil-opreators}) applied across regular or irregular grids. In this work, we consider regular three-dimensional grids that partition the space into cells, each of which with six neighbors. This regularity allows a cell to be addressed via a three-component index.

In stencil programs, optimizing individual stencils is often insufficient. Instead, chains of dependent stencils or entire programs must be optimized~\cite{gysi2015modesto}, e.g., using producer-consumer fusion to obtain maximum performance. We design the Stencil dialect to represent stencil programs consisting of stencil operations connected between them and with input/output data structures through data flow, with optional control flow (\autoref{sec-cf}), following the multi-level IR rewriting principles defined in \autoref{sec-mlrewriting}. Our dialect is explicitly decomposed (\textbf{P3}) into the high level, where we model the data flow between operators, and the low level, where we model the parallel execution of individual operators. The former enables flow rerouting transformations where a stencil operator can be seen as a unit (as opposed to lower-level IRs where a stencil is a collection of arithmetic instructions), while the latter supports parallelism exploitation (\textbf{P1}). The level separation also participates in progressive lowering (\textbf{P2}).

The dialect is not designed as a user-facing DSL, but as a compiler IR that supports transformations (\textbf{P1}\&\textbf{P3}) by (a) keeping the stencil concepts high-level so they can be moved as a unit, (b) imposing no specific execution order so as to expose parallelism, and (c) using value-semantics instead of allocating storage objects to avoid costly buffer analysis.



\begin{figure}
  \centering
  \includegraphics[width=\columnwidth]{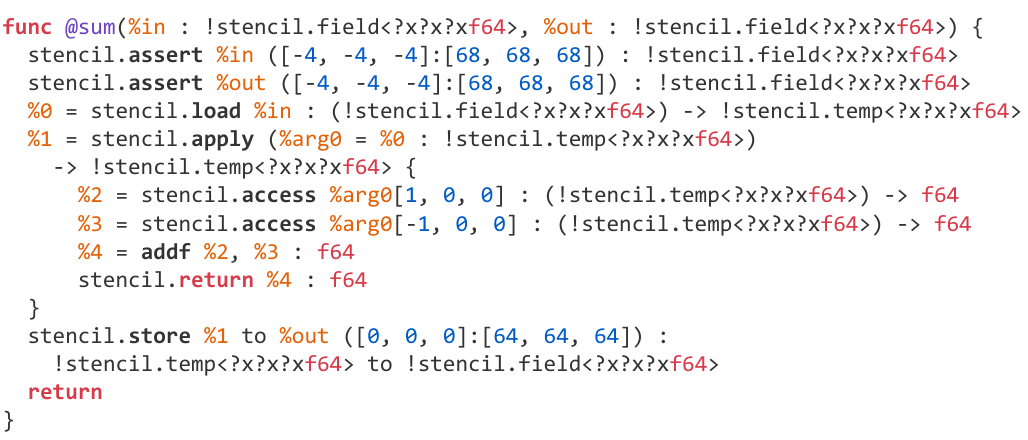}
  \caption{Example stencil program that evaluates a simple stencil on the array \textbf{\%in} and stores the result to the array \textbf{\%out}.}
  \label{fig-example}
\end{figure}

\subsection{Dialect Overview}
\label{sec-exampleprog}

The Stencil dialect focuses on concepts specific to stencils and relies on MLIR's Standard dialect to express the actual computation (\textbf{P2}). \autoref{fig-example} shows a stencil program that, for every point of a 64x64x64-element domain, adds the left and the right neighbor of the input array \textbf{\%in} and stores the result to the output array \textbf{\%out}. The ``stencil'' prefix identifies the operations and types from this dialect.

The dialect defines two types. A !stencil.\textbf{field} is a multi-dimensional array that stores an element for all points of the regular grid. Inputs and outputs of a stencil program have this type. A !stencil.\textbf{temp} is a multi-dimensional collection of elements on a hyper-rectangular subdomain of the regular grid. Temporaries have value semantics and are initially not backed by storage. Values of this type either point to a subdomain of an input array or keep the results computed by a stencil operator. Both types store single- or double-precision floating-point elements (f32 or f64) on a one-, two-, or three-dimensional domain.

The Stencil dialect also defines six operations. In the example, the stencil.\textbf{assert} operations specify the static shape of the arrays. The stencil.\textbf{load} operation takes the input array and returns a temporary that points to the input elements consumed by the subsequent stencil. The stencil.\textbf{apply} operation executes the stencil and defines a value that keeps the results of the computation. Its nested region implements the stencil operator for one point of the iteration domain. The stencil.\textbf{access} operations read the input temporary at a constant offset relative to the current position in the iteration domain, while the stencil.\textbf{return} operation sets the output at the current position. In between, we use the Standard dialect to sum the left and the right neighbor elements. The stencil.\textbf{store} operation finally stores the result of the computation to the output array. Its range attribute thereby specifies the domain written by the stencil program. Our compiler utilizes the output ranges to automatically infer the access ranges and iteration domains for the entire stencil program (cf.~\autoref{sec-preptrafos}).


\begin{figure}
  \centering
  \includegraphics[width=\columnwidth]{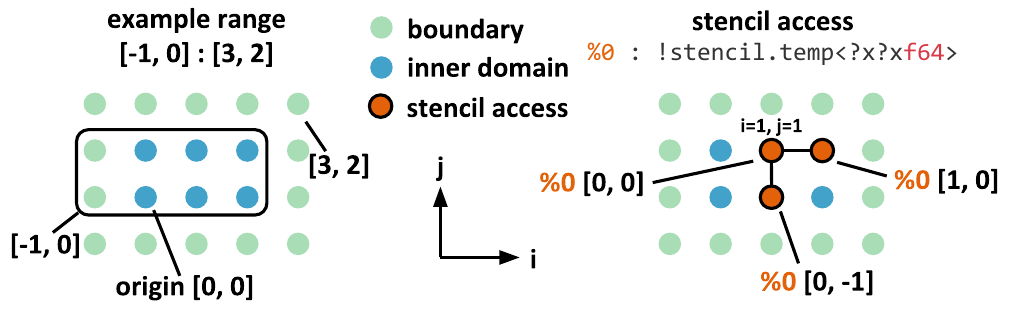}
  \caption{Example range (left) defined by an inclusive lower and an exclusive upper bound and stencil accesses (right) expressed relative to the current position (i=1, j=1).}
  \label{fig-ranges}
\end{figure}

\subsection{Shapes \& Domains}

The range notation is essential to specify stencil iteration domains and access ranges, especially given that stencils may be accessing inputs with indices that are outside of their computation domain, e.g., on boundaries.

\autoref{fig-ranges} shows our range notation (left) for a two-dimensional domain. The origin denotes the lower bound of the computation domain and has all coordinates set to zero. Ranges are specified given the absolute coordinates of an inclusive lower bound and an exclusive upper bound separated by a colon.

On GPUs, integer index computations are a significant performance bottleneck. As real-world stencil programs often execute stencils repeatedly for the same problem size, it is desirable for the stencil dialect to support size specialization for just-in-time compilation. It does so by defining storage shapes and iteration domains as numeric compile-time attributes (\textbf{P3}).

\subsection{Stencil Operators}
\label{sec-stencil-opreators}

A \emph{stencil operator} performs element-wise computations on all elements of a regular grid except for some constant-width boundary. It accesses the elements of input arrays at constant offsets relative to the coordinates of the output element. 

The stencil.\textbf{apply} operation contains a region that implements the stencil operator in terms of scalar operations on domain elements. The scalar operations are applied to all domain elements as in a loop nest. Stencil operator inputs and outputs correspond to values used and defined by this operation. Inputs are assumed to not alias, and element-wise computations are assumed to be independent (\textbf{P3}).
\grosser{They are known to not alias because we use an SSA graph}
\gysi{They could theoretically if the program inputs alias. It is acutally fine if the inputs alias, inputs and ouputs should not overlap?}

Individual elements of inputs are accessed using the stencil.\textbf{access} operation that reads an element at a constant offset. The lowering of the Stencil dialect later adds the constant access offset to the index of the current iteration (cf. \autoref{sec-stenlow}). \autoref{fig-ranges} shows the offset computation (right) for a two-dimensional stencil iteration domain. The region of the stencil operator must be terminated by a single stencil.\textbf{return} operation that accepts the value of the output element as an argument. Together, the stencil.\textbf{access} and stencil.\textbf{return} operations specify the memory access pattern of the stencil operator. Both of them are only valid as part of the stencil operator definition.



Real-world stencil programs from the weather and climate domain often implement dozens of dependent stencil operators. A stencil program thus needs additional means to orchestrate them.

\subsection{Stencil Programs}
\label{sec-stencil-programs}

A \emph{stencil program} executes a sequence of dependent stencil operators. It loads the data from the input arrays, implements the stencil operators inline, and stores the results to the output arrays. The SSA def-use graph of the program thus specifies the high-level data flow between the stencil operators (\textbf{P2}). Having both the high-level data flow and the inlined stencil operators in a single function facilitates code transformations across multiple stencil operators, eliminating any need for complex interprocedural analysis (\textbf{P1}).

Three additional operations are part of the program definition. The stencil.\textbf{assert} operation specifies the index range for an input or output array. A valid stencil program needs to define the index range for all input and output arrays. The stencil.\textbf{load} operation returns a temporary value that contains all input array elements accessed by dependent stencils. Conversely, the stencil.\textbf{store} operation stores the output of a stencil operator to the output array elements denoted by its range attribute.




\begin{figure}
  \centering
  \includegraphics[width=\columnwidth]{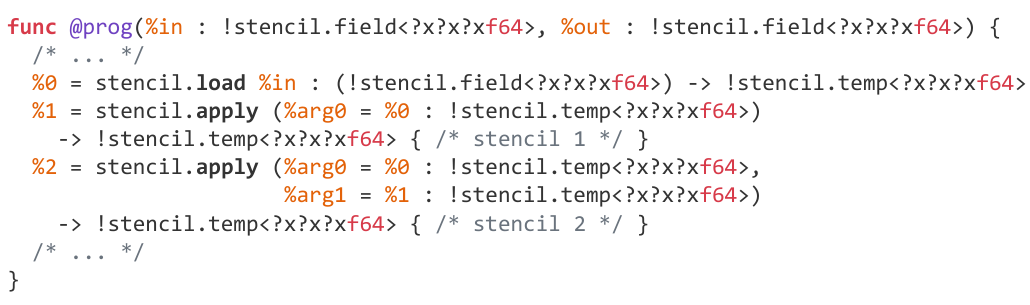}
  \caption{Stencil program that evaluates two dependent stencils.}
  \label{fig-multi}
\end{figure}

\autoref{fig-multi} shows a stencil program that executes two dependent stencils. The stencil.\textbf{load} operation returns a temporary holding the accessed \textbf{\%in} array elements. The second stencil operator uses the result of the first. In the end, the stencil.\textbf{store} operation stores the values computed by the second stencil to the \textbf{\%out} array.

All stencil program parameters have to be alias-free and are either loaded from or stored to as a unit. Intermediate results are kept in values of type !stencil.\textbf{temp} that are not initially backed by storage and are thus also alias-free. Given the value semantics of !stencil.\textbf{temp}, the def-use graph encodes the  data dependencies between the stencil operators (\textbf{P1\&P3}).



\subsection{Control Flow}
\label{sec-cf}

Real-world stencil applications are not limited to pure data flow semantics. We can use eager execution and just-in-time compilation to handle most of the control-flow at the program level and resort to MLIR's built-in Structured Control Flow (SCF) dialect to implement dynamic control flow inside the stencil operators.

\begin{figure}
  \centering
  \includegraphics[width=\columnwidth]{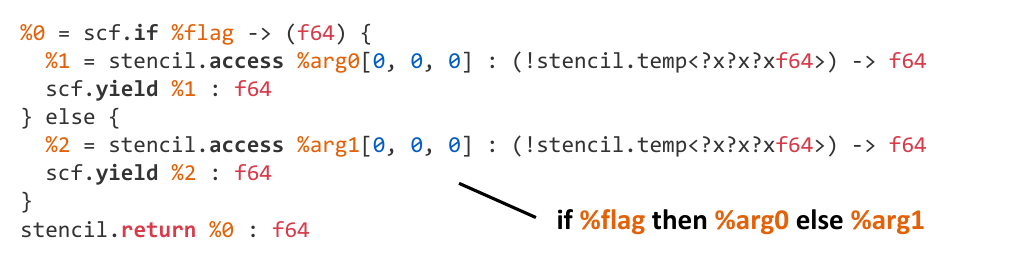}
  \caption{The SCF dialect enables the implementation of control flow inside the stencil operator.}
  \label{fig-ifelse}
\end{figure}

\autoref{fig-ifelse} shows a stencil that, depending on a flag, accesses one of two arguments. The scf.\textbf{if} operation conditionally executes either the ``then'' or the ``else'' region. In contrast to a regular if-else, the operation returns a result value that is set by the scf.\textbf{yield} operations. This representation makes the data flow explicit and maintains a single stencil.\textbf{return} operation per stencil. An alternative to the scf.\textbf{if} operation, is the \textbf{select} operation that chooses a value based on a condition. Supporting the scf.\textbf{if} operation requires no adaptation of our compiler (\textbf{P2}).

Our choice of the built-in MLIR SCF dialect exemplifies how progressive lowering, explicit separation, and composable abstractions enable reuse of compiler components in the multi-level IR rewriting scheme.

\section{Stencil Transformations}
\label{sec-stenciltrafos}

We distinguish three categories of transformations that work on the Stencil dialect: 1) performance optimizations, 2) transformations to prepare the lowering, and 3) the actual lowering.

\subsection{Optimizing Transformations}
\label{sec-opttrafos}

All optimizing transformations implemented for the Stencil dialect operate at a high-level and neither introduce explicit loops nor storage allocations (\textbf{P1}). 

\begin{figure}
  \centering
  \includegraphics[width=\columnwidth]{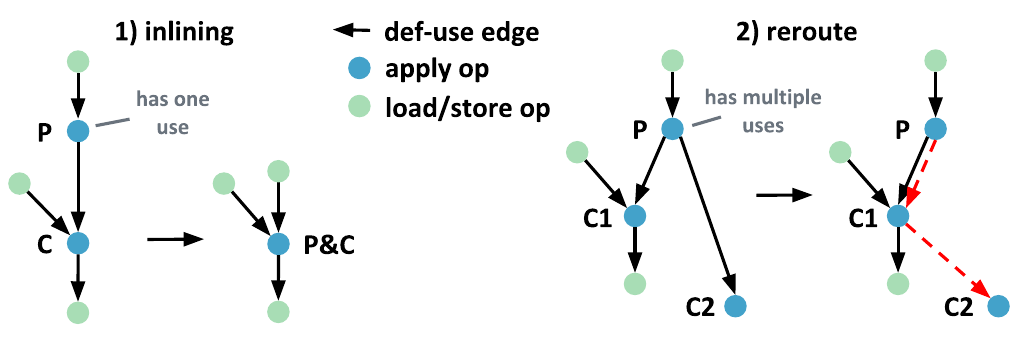}
  \caption{Two patterns enable the iterative producer-consumer fusion for entire stencil programs. The def-use edges represent the data flow between the stencil operations.}
  \label{fig-inlining}
\end{figure}

\smallskip
The \textbf{stencil inlining} pass applies fusion on the def-use graph of the stencil program. In particular, we repeatedly apply a stencil specific variant of producer-consumer fusion that replaces all accesses to producer results by inline computation. If the consumer accesses the producer at multiple offsets, we thus perform redundant computation for every point in the iteration domain. Inlining stencils in an arbitrary order may introduce circular dependencies. An input of the consumer may, for example, depend on another stencil that transitively depends on an output of the producer stencil. Instead of developing an algorithm to fuse the stencils in a valid order, we implement patterns that match and rewrite small subgraphs and use MLIR's greedy rewriter to apply them step-by-step.

\autoref{fig-inlining} shows our inlining patterns. The \emph{inlining} pattern matches a producer P and a consumer C if the producer has a single consumer. If the pattern matches, we remove the producer stencil and inline the computation into the consumer. Additionally, we update the argument and result lists of the fused stencils. The \emph{reroute} pattern matches a producer P and its consumers C1 to CN. If the pattern matches, we route all outputs of the producer through the consumer that executes next. The red (dashed) arrows mark the rerouted data dependencies. The former pattern implements the actual inlining, while the latter pattern prepares an inlining step.


Our inlining implementation introduces redundant computation even if the consumer accesses the same offset multiple times and always inlines the entire producer even if only one of its outputs is accessed. Dead code elimination and common subexpression elimination later clean up the code. These transformations rely on the stencil accesses being side-effect free (the stencil inputs are immutable and do not alias with the outputs). Our compiler currently implements no fusion heuristic and continuous inlining as long as one of the patterns applies.

\begin{figure}
  \centering
  \includegraphics[width=\columnwidth]{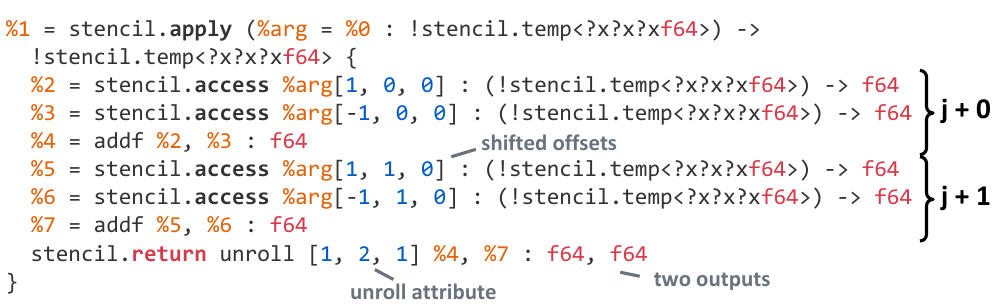}
  \caption{Unrolling two iterations of the example stencil along the j-dimension.}
  \label{fig-unrolling}
\end{figure}

\smallskip
The \textbf{stencil unrolling} pass replicates a stencil operator multiple times to update more than one grid point at once. \autoref{fig-unrolling} shows an unrolled version of our example program.

Unrolling is another example of a classical loop transformation implemented by our high-level dialect. Instead of transforming loops, our implementation annotates the high-level Stencil dialect and directly lowers to unrolled loops. In particular, we only modify the nested region attached to the stencil.\textbf{apply} operation but not its interface. Initially, we replicate the stencil computation once for every unrolled loop iteration and adjusts the access offsets. We also adapt the stencil.\textbf{return} operation to return the results of all unrolled loop iterations and annotate the unroll factor and dimension using an optional attribute.

The unrolling pass supports all unroll dimensions and unroll factors. Yet, the lowering is currently limited to unroll factors that divide the domain size evenly.

\smallskip
Inlining and unrolling improve the performance of stencil programs. Especially inlining reduces the off-chip data movement at the cost of introducing redundant computation. Unrolling can eliminate parts of the redundant computation since the unrolled stencil operator often evaluates the producer several times at the same offset. Instead of removing the redundant computation ourselves, we run the existing common subexpression elimination pass.

\subsection{Preparing the Lowering}
\label{sec-preptrafos}

After optimizing the stencil program, we infer all access ranges and iteration domains to prepare the lowering (\textbf{P2}).

\begin{figure*}
  \centering
  \includegraphics[width=\textwidth]{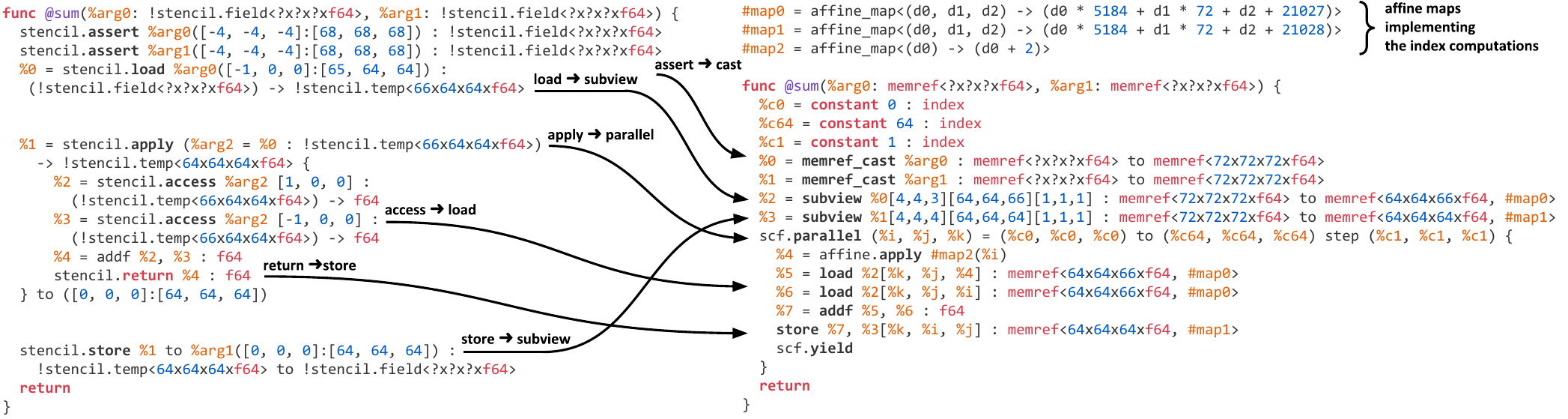}
  \caption{Conversion of the Stencil dialect to the MLIR SCF+Affine+Standard dialects that further lower to GPU abstractions.}
  \label{fig-lowering}
\end{figure*}

\begin{figure*}
  \centering
  \includegraphics[width=\textwidth]{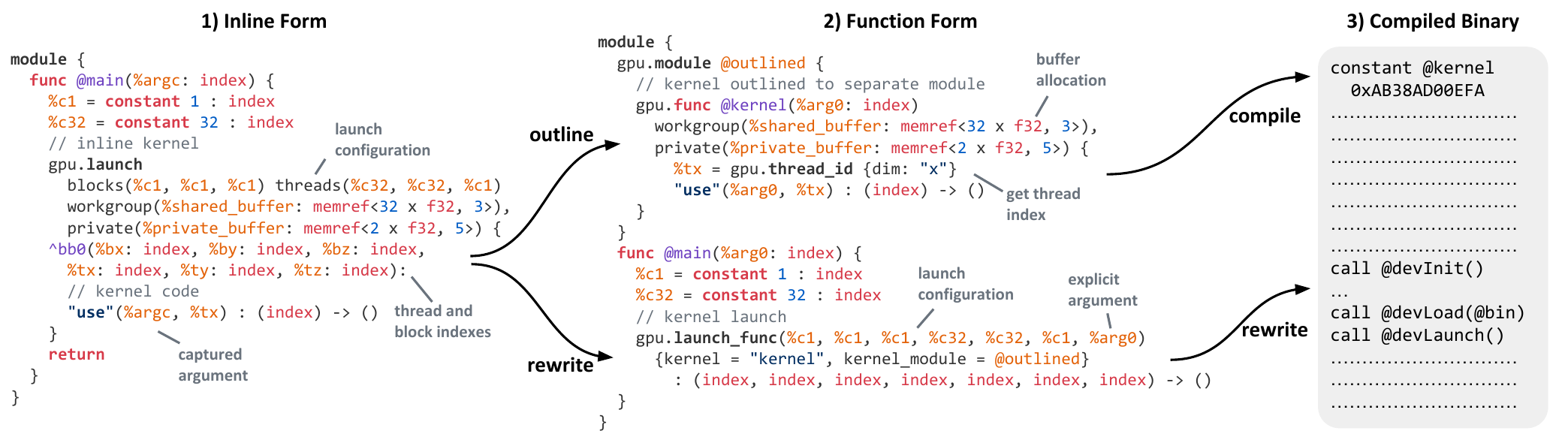}
  \caption{Lowering of an example kernel: 1) the inline form enables host device code motion and other transformations, 2) the function form isolates the device code in a distinct module that enables device-specific optimizations and separate host/device compilation, and 3) the binary embeds the kernel as constant data.}
  \label{fig-gpulowering}
\end{figure*}

\smallskip
The \textbf{shape inference} pass derives the access ranges for the input arrays and stencil operators of the program. It is necessary since a stencil program only defines the output ranges written by the program. The pass then starts from these output ranges and follows the use-def chains that define the dependencies of the stencil program and transitively extends the access ranges.

Our algorithm walks all operations of the stencil program in reverse order and annotates the access ranges using optional range attributes (\autoref{sec-stenlow} shows the lowering of the annotated example program). We compute these ranges as the minimal bounding box that contains all access extents of operations that consume a value defined by the current operation. If the consumer is a stencil.\textbf{load} operation, its access extent is equal to the output range attribute. If the consumer is a stencil.\textbf{apply} operation, the access extent is equal to the iteration domain extended by a minimal bounding box that contains all stencil accesses of the consumed values. Once the access range of a stencil.\textbf{load} operation is known, we also verify the input array is large enough.

Although the access extent analysis seemingly contradicts the progressive lowering idea (\textbf{P2}), it does not aim at recovering information that has been there before. Instead, it automates the error-prone manual access range specification.




\smallskip
Shape inference enables the lowering and has no performance impact.

\subsection{Lowering to Explicit Loops}
\label{sec-stenlow}

The stencil lowering applies conversion patterns to translate the individual stencil operations to their MLIR counterparts (\textbf{P2}). It is the last domain-specific part of our compilation pipeline, outlined in~\autoref{fig-overview}, that lowers our high-level stencil programs towards executable code.

Even at the Standard dialect level, MLIR provides rather high-level abstractions. The memref is a structured multi-dimensional buffer abstraction. It can have static or dynamic sizes, and an optional layout attribute defines the index computation if the layout diverges from the row-major format. This layout attribute also allows one to define strided hyper-rectangular views into a memory buffer, for example, with offsets and non-unit steps along each of the dimensions.  Another example is the scf.\textbf{parallel} operation that models a parallel multi-dimensional loop.

\autoref{fig-lowering} illustrates the lowering from the stencil dialect level to the MLIR Standard dialect level for the example introduced in \autoref{sec-exampleprog}. We define six conversion patterns that introduce explicit loops, index computations, memory accesses, and temporary storage. After this lowering, detecting stencil operators or access offsets requires analysis. Implementing domain-specific transformations consequently becomes harder. In turn, by introducing loops and temporary storage, we settle to program execution order but still maintain the parallel semantics needed for the subsequent GPU lowering (\textbf{P2}).

\section{The GPU Dialect}
\label{sec-gpulow}

Since GPUs remain a platform of choice to achieve high performance, we
construct our multi-level compiler to target these devices.
We designed following the principles defined in \autoref{sec-mlrewriting} and
implemented the GPU dialect for MLIR to this end with the goal of abstracting
the GPU execution model in a vendor-independent way.
In particular, it generalizes MLIR's NVVM, ROCm, and SPIR-V representations and
thus separates unified platform-independent device mapping (\textbf{P1}) from
platform-specific code generation (\textbf{P3}).
The GPU dialect is not intended as a generic SIMT execution model
(\textbf{P3}), nor as a raising target from lower-level abstractions
(\textbf{P2}).
The dialect exposes a set of GPU-specific concepts: hierarchical thread
structure (blocks, threads, warps); synchronization through barriers; memory
hierarchy (global, shared, private, constant memory); standard computational
primitives such as parallel reductions.
It is also designed to support separate host/device compilation in a single
module (\textbf{P3}).
The latter is made possible by MLIR modules recursively containing other
modules that can be processed differently.

\autoref{fig-gpulowering} shows the two forms of a kernel launch during the GPU lowering. The \emph{inline form} uses the gpu.\textbf{launch} operation to define the kernel inline. A nested region implements the kernel, and basic block arguments provide access to the thread and block identifiers. Explicit parameter handling is not needed since the values defined outside of the nested region remain visible. The \emph{function form} uses the gpu.\textbf{func} operation to implement the kernel as a separate function in a dedicated module launched by the gpu.\textbf{launch\_func} operation that represents the kernel invocation. Special operations provide access to the thread and block identifiers. All non-constant kernel arguments are passed in explicitly, while the constants are propagated into the kernel functions. Both the inline and the function form accept a GPU grid configuration and support the declarative allocation of buffers in the different levels of the GPU memory hierarchy. The kernel code expresses the computation for a single thread, following the SIMT model, and specialized mechanisms provide access to the thread and block identifiers. Thereby GPU-specific primitives such as barrier synchronization, shuffles, and ballots are only available inside a kernel launch.

Additionally, \autoref{fig-gpulowering} illustrates the main steps of the GPU lowering, starting from the inline form (left), through the function form (middle), down to the compiled binary (right). A parallel loop nest can be converted in-place to the inline form, using loop bounds as GPU grid configuration. After the conversion, we apply common subexpresison and dead code elimination, canonicalization and propagate constants inside the GPU kernel to minimize host/device memory traffic. Common SSA-based transformations apply seamlessly across the host/device boundary thanks to the kernel being inlined with no visibility restrictions. The kernel is then outlined into a separate function in a dedicated GPU module. Functions called by the kernel are copied into the module, and values defined outside the kernel are passed in as function arguments. This results in kernels living in a separate module to enable the separate host/device optimization and compilation. The kernel bodies are no longer visible to intra-procedural optimizations on the host code. The GPU module is finally converted through a dedicated dialect to a platform-specific representation (e.g., PTX), and using the vendor compiler (e.g., ptxas) compiles further down to a binary. The resulting binary is embedded as a global constant into the original module.  This approach enables, e.g., multi-versioning to support multiple architectures or kernel specialization for different-sized workloads. The original module extended with the binary constants then becomes a regular host module, that can be optimized, compiled, and executed. The kernel invocations thereby lower into calls to the device driver library or runtime environment.

\section{Evaluation}
\label{sec-eval}

We evaluate the Open Earth Compiler on real-world stencil programs derived from the most performance critical parts of the COSMO and FV3 climate models and compare its performance to state-of-the-art code generation frameworks.

\subsection{Experimental Setup}

We run our experiments on an NVIDIA Tesla V100-SXM2 with a memory bandwidth of 900 GB/s. We use the CUDA toolkit 10.1 with driver version 435.21 and benchmark two domain sizes 128x128x60 (small) and 256x256x60 (large) for single-precision (f32) and double-precision (f64) floating-point numbers. We further measure the execution times on an AMD Radeon RX 5700 with a memory bandwidth of 448 GB/s. Due to the low double-precision peak performance of the gaming-oriented hardware, we omit the double-precision results for the AMD system. For all benchmarks, we report the median runtime of 100 measurements, and red error bars show the quartile runtime to quantify the measurement error. We do not time the initial kernel executions to avoid startup overheads such as host to device data copies. Additionally, we use the nvprof profiler to collect detailed performance data on the NVIDIA system. We guarantee correctness by comparing the outputs of all optimized kernel variants to naive C versions and ensure the results are within a relative error of $10^{-5}$ for single-precision (f32) and $10^{-10}$ for double-precision (f64) floating-point numbers.


\begin{table}[t]
  \centering
  \footnotesize
  \begin{tabular}{@{}lcccccc@{}}
    \toprule
    \multirow{2}{*}{\textbf{Name}} & \multirow{2}{*}{\textbf{Dims}} & \textbf{Apply} & \textbf{Inputs/} & \textbf{Arith.} & \textbf{Access} & \textbf{Control} \\
                                   &                                & \textbf{Ops} & \textbf{Outputs} & \textbf{Ops}   & \textbf{Ops}  & \textbf{Flow}                                                                     \\
    \midrule
    \textbf{COSMO}                 &                                &                                    &                  &                &                                    &                                   \\

    hdiffsa                        & 2                              & 4                                  & 4 / 1            & 21             & 22                                 & min                               \\
    hdiffsmag                      & 2                              & 6                                  & 8 / 2            & 56             & 38                                 & min/max                           \\
    hadvuv                         & 2                              & 8                                  & 6 / 2            & 80             & 45                                 & if                                \\
    hadvuv5th                      & 2                              & 8                                  & 6 / 2            & 112            & 53                                 & if                                \\
    fastwavesuv                    & 3                              & 6                                  & 9 / 2            & 43             & 32                                 & -                              \\
    \textbf{FV3}                   &                                &                                    &                  &                &                                    &                                   \\
    p\_grad\_c                     & 3                              & 3                                  & 7 / 2            & 24             & 25                                 & -                              \\
    nh\_p\_grad                    & 3                              & 5                                  & 8 / 2            & 47             & 48                                 & -                              \\
    uvbke                          & 2                              & 2                                  & 4 / 2            & 12             & 12                                 & -                              \\
    fvtp2d\_qi                     & 2                              & 5                                  & 5 / 2            & 27             & 23                                 & if                                \\
    fvtp2d\_qj                     & 2                              & 8                                  & 6 / 3            & 49             & 39                                 & if                                \\
    fvtp2d\_flux                   & 2                              & 5                                  & 7 / 2            & 28             & 22                                 & if                                \\
    \bottomrule
  \end{tabular}
  \caption{Characteristics of our benchmarks.}
  \label{tab-cosmo-testset}
\end{table}

\subsection{Benchmark Kernels}

We evaluate our compiler for a set of representative benchmarks\footnote{\url{https://github.com/spcl/open-earth-benchmarks}} derived from the dynamical cores of two popular climate and weather models. COSMO~\cite{cosmoorg} is a regional numerical weather prediction model used by seven national weather services in Europe. FV3~\cite{fv3} is the dynamical core of the CM4 and GEOS-5 global climate models and the global weather prediction system of the US National Weather Service. The dominant algorithmic motif of both codes are stencil computations on regular grids. Both models implement dozens of stencil operators to perform the numerical forward integration in time. Due to the explicit time integration, most stencils are purely horizontal with bounded domains of dependence. Some use the Thomas algorithm to perform implicit integration in the vertical direction.



All our benchmarks are part of the explicit integration. The hadvuv and hadvuv5th kernels implement third- and fifth-order horizontal advection, while the fvtp2d kernels implement a monotone two-dimensional finite volume advection operator. The hdiffsa and hdiffsmag kernels perform horizontal diffusion. The fastwavesuv kernel contains parts of the sound wave forward integration, while the p\_grad\_c and nh\_p\_grad kernels compute the three-dimensional pressure gradient. Finally, the uvbke kernel is a preprocessing step for the kinetic energy computation in FV3. 

Each benchmark executes an entire stencil program consisting of multiple stencil operators being applied on the three-dimensional domain. The stencil operators have different dimensionality (from one- to three-dimensional), have different width (two- to five-point), and some of them contain dynamic control flow. \autoref{tab-cosmo-testset} list core characteristics of our benchmarks such as the dimensionality of the stencil access patterns or the number of stencil operators, input/output arrays, arithmetic operations, and stencil.\textbf{access} operations. We observe that all kernels have a low arithmetic intensity (arithmetic operations per memory access), which explains why our compiler focuses on transformations to increase the data-locality.\gysi{discuss performance on different problem sizes}

\begin{figure}
  \centering
  \includegraphics[trim=0 0 0 32, clip, width=\columnwidth]{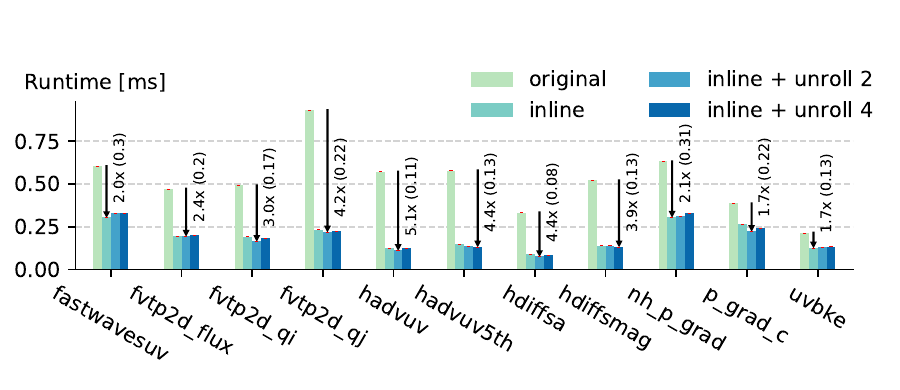}
  \includegraphics[trim=0 0 0 32, clip,  width=\columnwidth]{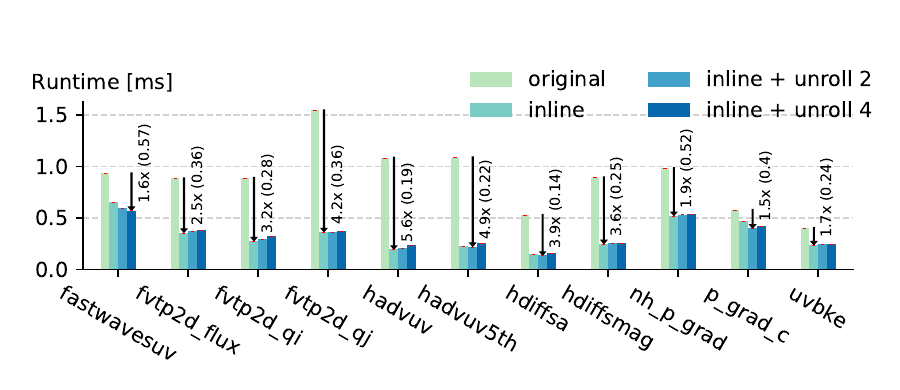}
  \caption{Runtimes at different optimization levels for f32 (top) and f64 (bottom) floating-point values for 256x256x60 on the NVIDIA system.}
  \label{fig-optimization-comparison}
\end{figure}

\begin{figure}
  \centering
  \includegraphics[trim=0 0 0 32, clip, width=\columnwidth]{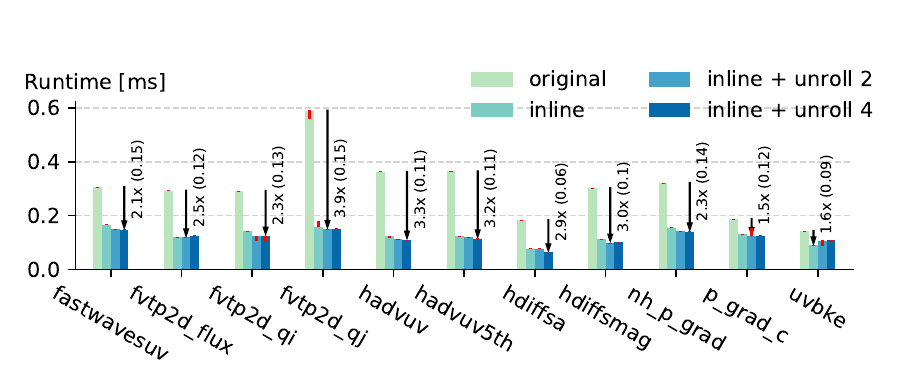}
  \includegraphics[trim=0 0 0 32, clip, width=\columnwidth]{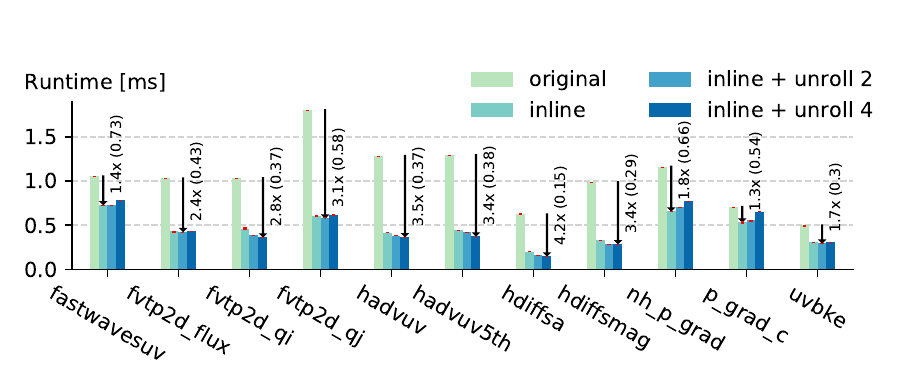}
    \caption{Runtimes at different optimization levels for f32 floating-point values for 128x128x60 (top) and 256x256x60 (bottom) on the AMD system.}
  \label{fig-amd-comparison}
\end{figure}

\subsection{Effectiveness of our Code Transformations}
\label{sec-effectiveness-of-trafos}

We first evaluate the effectiveness of the code transformations discussed in~\autoref{sec-opttrafos}. In total, we compare different optimization levels: 1) \emph{original}, 2) \emph{inline}, 3) \emph{inline+unroll(2)}, and 4) \emph{inline+unroll(4)}. Optimization level one applies no optimizing transformations (cf.~\autoref{sec-opttrafos}). Starting from optimization level two we apply stencil inlining, and the optimization levels three and four additionally perform stencil unrolling by factor two and four, respectively.

\autoref{fig-optimization-comparison} and \autoref{fig-amd-comparison} compare the runtime for all benchmarks at different optimization levels on the NVIDIA and the AMD system, respectively. We choose these platforms to demonstrate our device-specific code generation for two entirely different instruction set architectures and runtime environments. We show data for both problem sizes (AMD) and f32 and f64 arithmetic (NVIDIA) and observe significant speedups for stencil inlining independent of benchmark and system. In comparison, stencil unrolling has a smaller effect and sometimes is even detrimental to performance. In the plots, we annotate the speedup and the runtime of the best performing version.

\begin{figure}
  \centering
  \includegraphics[trim=0 0 0 25, clip,width=\columnwidth]{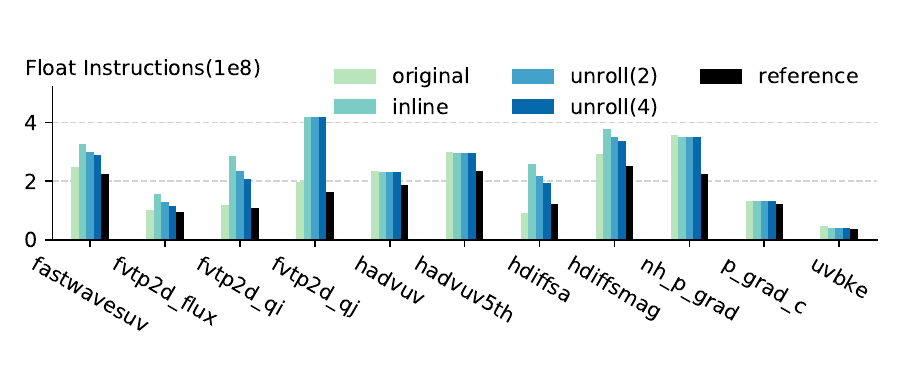}
  \includegraphics[trim=0 0 0 25, clip,width=\columnwidth]{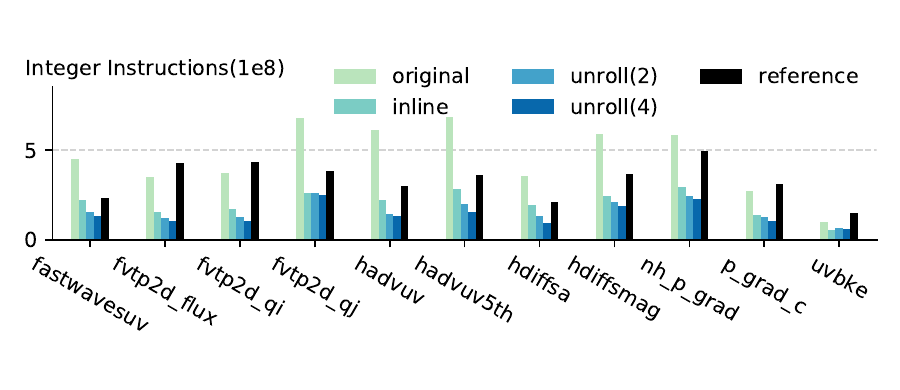}
  \includegraphics[trim=0 0 0 25, clip,width=\columnwidth]{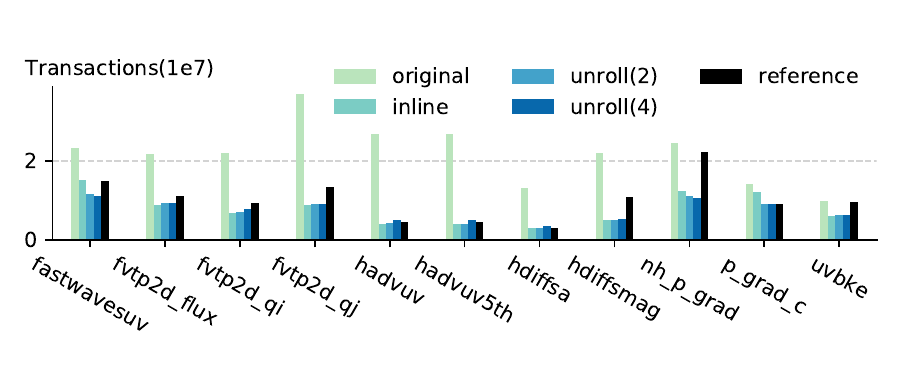}
    \caption{Number of float instructions (top), integer instructions (middle), and the device memory transactions (bottom) executed at different optimization levels in comparison to COSMO and FV3 reference implementations (cf.~\autoref{sec-state-of-the-art}) for f64 floating-point values and 256x256x60 (these measurements are profiling results collected using nvprof).}
  \label{fig-instr-trans}
\end{figure}

\begin{figure}
  \centering
  \includegraphics[trim=0 0 0 25, clip,width=\columnwidth]{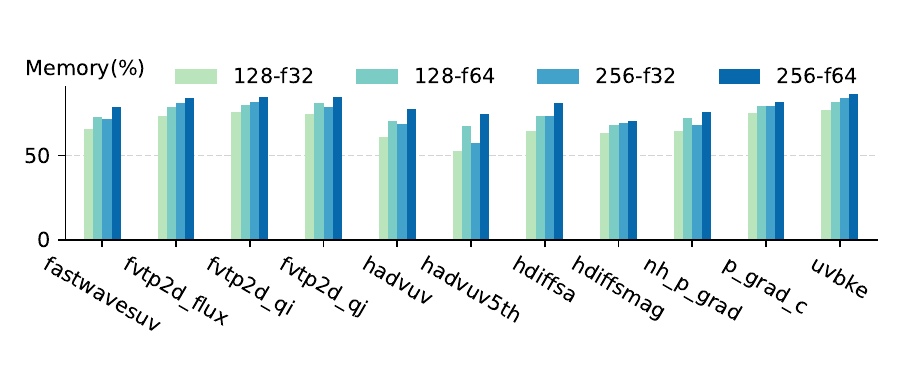}
  \includegraphics[trim=0 0 0 25, clip,width=\columnwidth]{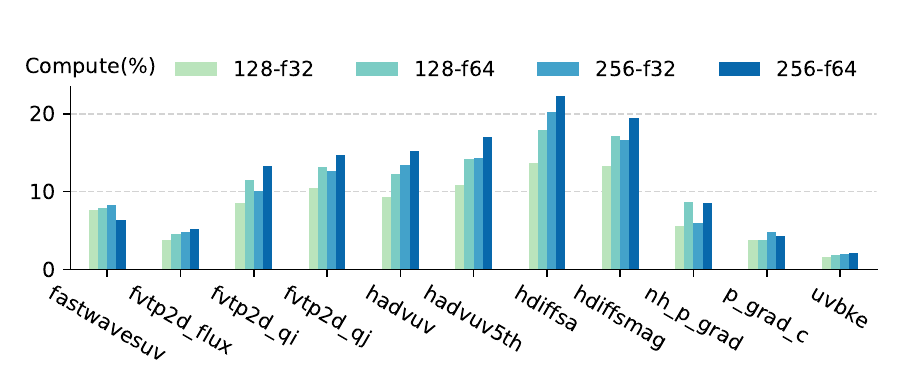}
  \caption{Utilization of the peak compute throughput (top) and the peak memory bandwidth (bottom) for the best performing kernel variants in percent (these measurements are profiling results collected using nvprof).}
  \label{fig-cmp-profile}
\end{figure}

Inlining all stencil operators eliminates the accesses of temporary buffers, and the program inputs are loaded precisely once. We thus expect stencil inlining to have a strong effect on performance due to the resulting bandwidth reduction. \autoref{fig-instr-trans} confirms that stencil inlining reduces the data movement between the device memory and the L2 cache (bottom). At the same time, inlining introduces redundant computation. \autoref{fig-instr-trans} quantifies this extra computation (top) but also shows that stencil inlining eliminates parts of the index computation (middle). Overall the bandwidth reduction and the eliminated index computation overcompensate the additional redundant computation, especially considering the low arithmetic intensity of our kernels.

Stencil unrolling removes redundant computations and improves the data-locality in case the data accesses of the unrolled loop iterations overlap. \autoref{fig-instr-trans} confirms that stencil unrolling reduces both redundant computation (top) and index computation (middle). We also observe less device memory transactions (bottom) for the three-dimensional fastwavesuv, nh\_p\_grad, and p\_grad\_c kernels due to improved data-locality. At the same time, stencil unrolling increases the register pressure and reduces the available parallelism. As a consequence, we do not expect all stencil programs to benefit from stencil unrolling. Instead, the optimization's effectiveness depends on the complexity of the stencil program (register pressure), the available parallelism, and the potential bandwidth reduction.

\autoref{fig-cmp-profile} illustrates the memory bandwidths and compute throughputs achieved by the best performing kernel versions. We observe very high memory bandwidth utilization up to 86\% and low compute utilization below 22\%. These results demonstrate the importance of aggressive stencil inlining and confirm the redundant computation is less of a concern.

In summary, we show that our code transformations yield significant speedups. We attain comparable speedups and bandwidth rectified performance on both the NIVIDA and the AMD system despite their entirely different instruction set architecture and runtime environment, demonstrating the effectiveness of our vendor-independent GPU execution model introduced in~\autoref{sec-gpulow}. Selecting the optimal unroll factor or finding good fusion choices for a specific benchmark and target system combination is not the scope of our work. Instead, we employ empirical tuning to find the best unroll factor and always fuse all stencil operators (optimizing larger stencil programs will require a fusion heuristic).

\muellch{Switch between int32 and int64, Plot: best performance (int32) vs best performance (int64)}

\subsection{Comparison to State-of-The-Art Solutions}
\label{sec-state-of-the-art}

We now compare the runtime of the kernels optimized by our compiler to Stella~\cite{gysi2015stella} and Dawn~\cite{osuna2020dawn} generated CUDA~\cite{cuda2008} implementations. Stella is a C++ embedded DSL used to run the GPU version of COSMO in production. Dawn is a research compiler that lowers high-level stencil programs to efficient CUDA code. Both implement overlapped tiling~\cite{holewinski2012overlapped} using shared memory and stream data~\cite{rawat2016streaming} in registers along the k-dimension. This execution model limits the redundant computation to the tile boundaries. In comparison, our stencil inlining plus unrolling performs additional redundant computation at the thread level but requires no thread synchronization during the kernel execution.

\begin{figure}
  \centering
  \includegraphics[trim=0 0 0 25, clip,width=\columnwidth]{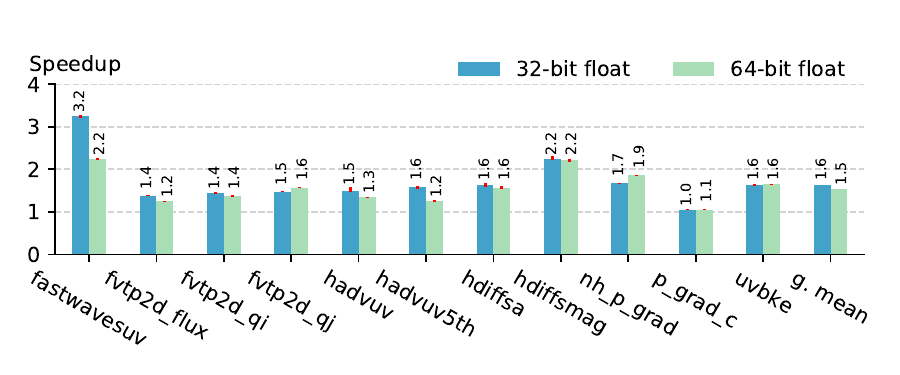}
  \includegraphics[trim=0 0 0 25, clip,width=\columnwidth]{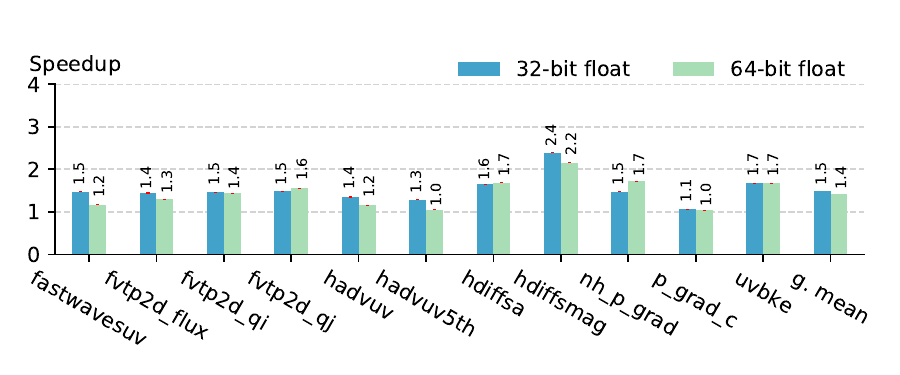}
  \caption{Speedup of our compiler over Stella (COSMO) and Dawn (FV3) for 128x128x60 (top) and 256x256x60 (bottom) on the NVIDIA system.}
  \label{fig-cmp-stella}
\end{figure}

\autoref{fig-cmp-stella} \wicky{is it hard to add the reference implementation to the plots in the appendix? For the sake of readability I can see omitting confidence intervals in the main plots. Are we sure we don't want them in the appendix either?} compares for all benchmarks the best performing variant generated by the Open Earth Compiler to the Stella (COSMO) and Dawn (FV3) counterparts. We outperform the state-of-the-art and obtain speedups that range from 1.4x to 1.6x depending on problem size and precision. We thus perform better for all configurations especially for the smaller problem size and single-precision arithmetic. The speedups are similar, with the exception of the fastwavesuv kernel. Stella implements the fastwavesuv kernel using a sequential loop in the k-dimension, which results in suboptimal performance for the smaller problem sizes.

We attribute the performance of our compiler to the simple execution model. It fuses all stencil operators and stores temporaries in registers to limit the data movement, and it introduces redundant computation instead of thread synchronizations to avoid parallelization overheads. Its only disadvantage is the redundant computation, which due to the low arithmetic intensity of our kernels shown in~\autoref{fig-cmp-profile}, is less critical. In \autoref{fig-instr-trans}, we indeed observe lower device memory bandwidth requirements (bottom) and notable amounts of redundant computation (top) compared to Stella (COSMO) and Dawn (FV3) reference implementations. Having compile-time information about storage shapes and iteration domains in return eliminates parts of the index computation (middle). Avoiding synchronizations at the cost of additional computation and leveraging compile-time information thus turn out to be beneficial compared to the Stella and Dawn execution models.

Our compiler, despite its simplicity, outperforms Stella and Dawn on raw stencil programs. The results demonstrate the quality of our code generation and the potential of stencil inlining plus unrolling compared to overlapped tiling and streaming, the standard solution in the field.

\begin{figure}
  \centering  
  \includegraphics[width=\columnwidth]{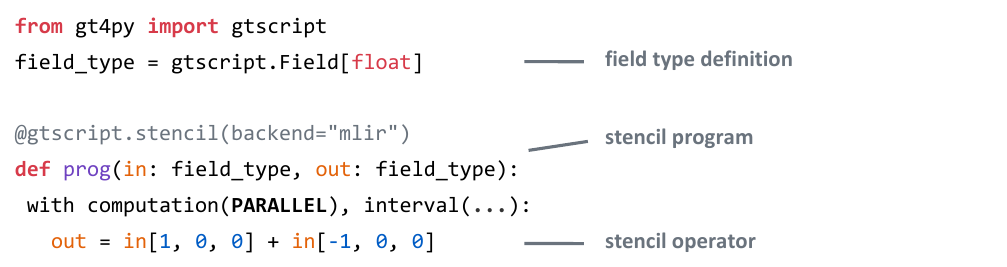}
  \caption{GT4Py version of the example stencil program.}
  \label{fig-gt4py}
\end{figure}

\subsection{Lowering User-facing Code to the Stencil Dialect}

We design our Stencil dialect as a target for user-facing domain-specific languages. To study the feasibility of such a lowering, we integrate our compiler with GridTools for Python (GT4Py)~\cite{gt4pygithub}, an embedded domain-specific language for weather and climate.

In \autoref{fig-gt4py}, we show a GT4Py version of the example stencil program introduced in~\autoref{sec-exampleprog}. GT4Py lowers the user code to an internal IR, consisting of a compute domain, data field descriptors, and a set of computations in the form of an abstract syntax tree (AST). We traverse the AST to emit MLIR, compile the resulting MLIR program to a binary, and produce Python bindings to link the generated binary to the calling program.

The functional lowering demonstrates our compiler's applicability in the context of an end-to-end solution for the weather and climate domain.

\balance
\section{Related Work}

Accelerated systems made programming model innovations inevitable. Kokkos~\cite{edwards2013kokkos} and Raja~\cite{rajagithub} are C++ performance portability layers. PENCIL~\cite{baghdadi2015pencil} and Polly-ACC~\cite{grosser2016pollyacc} automate the accelerator mapping using the polyhedral model. DaCe~\cite{bennun2019dace} allows performance engineers to select and develop target-specific transformations. All approaches are generic and, for the same level of performance and automation, solve a more complex problem than a domain-specific compiler.

Machine learning today drives the development of domain-specific compilers~\cite{leary2017xla,tianqi2018tvm}. The development of stencil compilers started even earlier: Halide~\cite{ragan2013halide} and Polymage~\cite{mullapudi2015polymage} tune image processing pipelines, Pochoir~\cite{tang2011pochoir} implements cache-oblivious tiling, SDSLc~\cite{rawat2015SDSLc} supports many targets (SIMD, GPU, and FPGA), Panda~\cite{sourouri2017panda} supports distributed memory, and YASK~\cite{yask2016} specifically targets Intel processors. Lift~\cite{hagedorn2018liftstencils} has also been shown effective for stencil codes. The variety of different solutions demonstrates the importance of a shared compiler infrastructure.

Multiple projects work on solutions for weather and climate. The CLIMA~\cite{climagithub} effort develops a novel earth system model using the Julia language. The LFRic~\cite{adams2019lfric} climate modeling system relies on the Python-based PSyclone compiler. Stella~\cite{gysi2015stella} and GridTools~\cite{gridtoolsgithub} use C++ template metaprogramming to support CPU and GPU systems. CLAW~\cite{clement2018claw} and Hybrid Fortran~\cite{mueller2018hybridfortran} extend Fortran to achieve performance portability. Despite their heterogeneity, all of these approaches could benefit from a shared compiler infrastructure.

Several frameworks support the development of domain-specific compilers. AnyDSL~\cite{leissa2018anydsl} supports partial evaluation using minimal annotations in the Impala front end language. Lightweight modular staging~\cite{rompf2010staging} is a technique that uses Scala's type system to transform codes before their execution. It forms the basis of the Delite~\cite{sujeeth2014delite} compiler framework. Lua script similarly supports staging via the Terra~\cite{devito2013terra} low-level language. Lift~\cite{steuwer2017lift} finally combines a functional language and rewrite rules to generate performance portable code. MLIR~\cite{lattner2020mlir} is the only full-fledged compiler infrastructure among these contenders, not limited in terms of optimizations, and not tied to a particular front end language.

Stencil optimizations for GPU targets are a well-researched topic. Tiling~\cite{nguyen2010blocking,holewinski2012overlapped,maruyama2014optimizing,grosser2014hybrid,grosser2014split,matsumura2020an5d} and fusion~\cite{rawat2018fusion,gysi2015modesto,wahib2014fusion} are the core optimizations for bandwidth-limited low-order stencils as they appear in weather and climate. Other works optimize the resource utilization~\cite{rawat2018regopt,rawat2016streaming} or discuss the optimization of high-order stencils~\cite{zhao2019stencil,rawat2015SDSLc}. Our compiler implements a variant of overlapped tiling~\cite{holewinski2012overlapped} that introduces redundant computation for every thread.

\section{Conclusion}

We introduced multi-level IR rewriting, an approach to building reusable components for domain-specific compilers. This approach is illustrated through the design and implementation of the Open Earth Compiler, which provides a high-performance compilation flow for weather and climate modeling. We demonstrated that thanks to multi-level IR rewriting, a small yet self-consistent set of high-level operations specifically designed for stencil computations is sufficient to achieve better performance than state-of-the-art DSL compilers. Contrary to the latter, the Open Earth Compiler relies on existing and new reusable compiler abstractions, including the GPU kernel abstraction we introduced, by decoupling domain-specific and target-specific code transformations. Our evaluation of eleven stencil programs relevant to existing climate models, COSMO and FV3, demonstrates that the Open Earth Compiler generates code that is up to 3.2x faster than state-of-the-art solutions and delivers a geomean speedup between 1.4x and 1.6x across problem sizes and precisions. We suggest that multi-level IR rewriting and the associated design principles is a promising approach to rapidly design and deploy domain-specific compilers that can take advantage of reusable components of the MLIR ecosystem.

\section*{Acknowledgements}
We thank Jean-Michel Gorius for his foundational stencil compiler work and the continuous support of our project. We appreciate the assistance of Carlos Osuna and the MeteoSwiss compiler team throughout the project. Finally, we thank Hannes Vogt for providing the initial out-of-tree development setup and Felix Thaler for sharing performance results that inspired our work.
This project has received funding from a Huawei donation and through the Swiss National Science Foundation under the Ambizione programme (grant PZ00P2168016) as well as ARM Holdings plc and Xilinx Inc in the context of Polly Labs.


\clearpage
\bibliographystyle{IEEEtran}
\bibliography{references}

\begin{thebibliography}{10}
\providecommand{\url}[1]{#1}
\csname url@samestyle\endcsname
\providecommand{\newblock}{\relax}
\providecommand{\bibinfo}[2]{#2}
\providecommand{\BIBentrySTDinterwordspacing}{\spaceskip=0pt\relax}
\providecommand{\BIBentryALTinterwordstretchfactor}{4}
\providecommand{\BIBentryALTinterwordspacing}{\spaceskip=\fontdimen2\font plus
\BIBentryALTinterwordstretchfactor\fontdimen3\font minus
  \fontdimen4\font\relax}
\providecommand{\BIBforeignlanguage}[2]{{%
\expandafter\ifx\csname l@#1\endcsname\relax
\typeout{** WARNING: IEEEtran.bst: No hyphenation pattern has been}%
\typeout{** loaded for the language `#1'. Using the pattern for}%
\typeout{** the default language instead.}%
\else
\language=\csname l@#1\endcsname
\fi
#2}}
\providecommand{\BIBdecl}{\relax}
\BIBdecl

\bibitem{mullapudi2015polymage}
\BIBentryALTinterwordspacing
R.~T. Mullapudi, V.~Vasista, and U.~Bondhugula, ``{PolyMage}: Automatic
  optimization for image processing pipelines,'' in \emph{Proceedings of the
  Twentieth International Conference on Architectural Support for Programming
  Languages and Operating Systems}, ser. ASPLOS '15.\hskip 1em plus 0.5em minus
  0.4em\relax New York, NY, USA: Association for Computing Machinery, 2015, pp.
  429--443. [Online]. Available: \url{https://doi.org/10.1145/2694344.2694364}
\BIBentrySTDinterwordspacing

\bibitem{tianqi2018tvm}
\BIBentryALTinterwordspacing
T.~Chen, T.~Moreau, Z.~Jiang, L.~Zheng, E.~Yan, H.~Shen, M.~Cowan, L.~Wang,
  Y.~Hu, L.~Ceze, C.~Guestrin, and A.~Krishnamurthy, ``{TVM}: An automated
  end-to-end optimizing compiler for deep learning,'' in \emph{13th {USENIX}
  Symposium on Operating Systems Design and Implementation ({OSDI} 18)}.\hskip
  1em plus 0.5em minus 0.4em\relax Carlsbad, CA: {USENIX} Association, Oct.
  2018, pp. 578--594. [Online]. Available:
  \url{https://www.usenix.org/conference/osdi18/presentation/chen}
\BIBentrySTDinterwordspacing

\bibitem{rawat2015SDSLc}
\BIBentryALTinterwordspacing
P.~Rawat, M.~Kong, T.~Henretty, J.~Holewinski, K.~Stock, L.-N. Pouchet,
  J.~Ramanujam, A.~Rountev, and P.~Sadayappan, ``{SDSLc}: A multi-target
  domain-specific compiler for stencil computations,'' in \emph{Proceedings of
  the 5th International Workshop on Domain-Specific Languages and High-Level
  Frameworks for High Performance Computing}, ser. WOLFHPC '15.\hskip 1em plus
  0.5em minus 0.4em\relax New York, NY, USA: Association for Computing
  Machinery, 2015. [Online]. Available:
  \url{https://doi.org/10.1145/2830018.2830025}
\BIBentrySTDinterwordspacing

\bibitem{yask2016}
C.~{Yount}, J.~{Tobin}, A.~{Breuer}, and A.~{Duran}, ``{YASK}—yet another
  stencil kernel: A framework for hpc stencil code-generation and tuning,'' in
  \emph{2016 Sixth International Workshop on Domain-Specific Languages and
  High-Level Frameworks for High Performance Computing (WOLFHPC)}, 2016, pp.
  30--39.

\bibitem{sourouri2017panda}
\BIBentryALTinterwordspacing
M.~Sourouri, S.~B. Baden, and X.~Cai, ``Panda: A compiler framework for
  concurrent {CPU+GPU} execution of 3d stencil computations on
  {GPU}-accelerated supercomputers,'' \emph{Int. J. Parallel Program.},
  vol.~45, no.~3, pp. 711--729, Jun. 2017. [Online]. Available:
  \url{https://doi.org/10.1007/s10766-016-0454-1}
\BIBentrySTDinterwordspacing

\bibitem{tang2011pochoir}
\BIBentryALTinterwordspacing
Y.~Tang, R.~A. Chowdhury, B.~C. Kuszmaul, C.-K. Luk, and C.~E. Leiserson, ``The
  pochoir stencil compiler,'' in \emph{Proceedings of the Twenty-Third Annual
  ACM Symposium on Parallelism in Algorithms and Architectures}, ser. SPAA
  '11.\hskip 1em plus 0.5em minus 0.4em\relax New York, NY, USA: Association
  for Computing Machinery, 2011, pp. 117--128. [Online]. Available:
  \url{https://doi.org/10.1145/1989493.1989508}
\BIBentrySTDinterwordspacing

\bibitem{ragan2013halide}
\BIBentryALTinterwordspacing
J.~Ragan-Kelley, C.~Barnes, A.~Adams, S.~Paris, F.~Durand, and S.~Amarasinghe,
  ``Halide: A language and compiler for optimizing parallelism, locality, and
  recomputation in image processing pipelines,'' in \emph{Proceedings of the
  34th ACM SIGPLAN Conference on Programming Language Design and
  Implementation}, ser. PLDI '13.\hskip 1em plus 0.5em minus 0.4em\relax New
  York, NY, USA: Association for Computing Machinery, 2013, pp. 519--530.
  [Online]. Available: \url{https://doi.org/10.1145/2491956.2462176}
\BIBentrySTDinterwordspacing

\bibitem{leary2017xla}
C.~Leary and T.~Wang, ``{XLA}: Tensorflow, compiled,'' \emph{TensorFlow Dev
  Summit}, 2017.

\bibitem{gysi2015stella}
\BIBentryALTinterwordspacing
T.~Gysi, C.~Osuna, O.~Fuhrer, M.~Bianco, and T.~C. Schulthess, ``{STELLA}: A
  domain-specific tool for structured grid methods in weather and climate
  models,'' in \emph{Proceedings of the International Conference for High
  Performance Computing, Networking, Storage and Analysis}, ser. SC '15.\hskip
  1em plus 0.5em minus 0.4em\relax New York, NY, USA: Association for Computing
  Machinery, 2015. [Online]. Available:
  \url{https://doi.org/10.1145/2807591.2807627}
\BIBentrySTDinterwordspacing

\bibitem{fuhrer2014stella}
\BIBentryALTinterwordspacing
O.~Fuhrer, C.~Osuna, X.~Lapillonne, T.~Gysi, B.~Cumming, M.~Bianco, A.~Arteaga,
  and T.~Schulthess, ``Towards a performance portable, architecture agnostic
  implementation strategy for weather and climate models,''
  \emph{Supercomputing Frontiers and Innovations}, vol.~1, no.~1, 2014.
  [Online]. Available: \url{https://superfri.org/superfri/article/view/17}
\BIBentrySTDinterwordspacing

\bibitem{quinlan2000rose}
D.~Quinlan, ``{ROSE}: Compiler support for object-oriented frameworks,''
  \emph{Parallel Processing Letters}, vol.~10, no. 02n03, pp. 215--226, 2000.

\bibitem{lattner2004llvm}
C.~Lattner and V.~Adve, ``{LLVM}: A compilation framework for lifelong program
  analysis \& transformation,'' in \emph{International Symposium on Code
  Generation and Optimization, 2004. CGO 2004.}\hskip 1em plus 0.5em minus
  0.4em\relax IEEE, 2004, pp. 75--86.

\bibitem{seismic}
G.~A. {McMechan}, ``{Migration by Extrapolation of Time-Dependent Boundary
  {VALUES}},'' \emph{Geophysical Prospecting}, vol.~31, pp. 413--420, Jun.
  1983.

\bibitem{harris2013fv3dycore}
\BIBentryALTinterwordspacing
L.~M. Harris and S.-J. Lin, ``A two-way nested global-regional dynamical core
  on the cubed-sphere grid,'' \emph{Monthly Weather Review}, vol. 141, no.~1,
  pp. 283--306, 2013. [Online]. Available:
  \url{https://doi.org/10.1175/MWR-D-11-00201.1}
\BIBentrySTDinterwordspacing

\bibitem{cosmo}
M.~Baldauf, A.~Seifert, J.~F\"orstner, D.~Majewski, M.~Raschendorfer, and
  T.~Reinhardt, ``Operational convective-scale numerical weather prediction
  with the {COSMO} model: Description and sensitivities,'' \emph{Monthly
  Weather Review}, vol. 139, no.~12, pp. 3887--3905, 2011.

\bibitem{sujeeth2014delite}
\BIBentryALTinterwordspacing
A.~K. Sujeeth, K.~J. Brown, H.~Lee, T.~Rompf, H.~Chafi, M.~Odersky, and
  K.~Olukotun, ``Delite: A compiler architecture for performance-oriented
  embedded domain-specific languages,'' \emph{ACM Trans. Embed. Comput. Syst.},
  vol.~13, no.~4s, Apr. 2014. [Online]. Available:
  \url{https://doi.org/10.1145/2584665}
\BIBentrySTDinterwordspacing

\bibitem{rosen1988global}
B.~K. Rosen, M.~N. Wegman, and F.~K. Zadeck, ``Global value numbers and
  redundant computations,'' in \emph{Proceedings of the 15th ACM SIGPLAN-SIGACT
  symposium on Principles of programming languages}, 1988, pp. 12--27.

\bibitem{muchnick1998compilers}
S.~S. Muchnick, \emph{Advanced Compiler Design and Implementation}.\hskip 1em
  plus 0.5em minus 0.4em\relax San Francisco, CA, USA: Morgan Kaufmann
  Publishers Inc., 1998.

\bibitem{mckeeman1965peephole}
W.~M. McKeeman, ``Peephole optimization,'' \emph{Communications of the ACM},
  vol.~8, no.~7, pp. 443--444, 1965.

\bibitem{lattner2020mlir}
\BIBentryALTinterwordspacing
C.~Lattner, M.~Amini, U.~Bondhugula, A.~Cohen, A.~Davis, J.~Pienaar, R.~Riddle,
  T.~Shpeisman, N.~Vasilache, and O.~Zinenko, ``{MLIR}: A compiler
  infrastructure for the end of moore's law,'' 2020, {CoRR} preprint
  arXiv:2002.11054v2 [cs.PL]. [Online]. Available:
  \url{https://arxiv.org/abs/2002.11054v2}
\BIBentrySTDinterwordspacing

\bibitem{grosser2016pollyacc}
\BIBentryALTinterwordspacing
T.~Grosser and T.~Hoefler, ``{Polly-ACC} transparent compilation to
  heterogeneous hardware,'' in \emph{Proceedings of the 2016 International
  Conference on Supercomputing}, ser. ICS '16.\hskip 1em plus 0.5em minus
  0.4em\relax New York, NY, USA: Association for Computing Machinery, 2016.
  [Online]. Available: \url{https://doi.org/10.1145/2925426.2926286}
\BIBentrySTDinterwordspacing

\bibitem{verdoolaege2013ppcg}
\BIBentryALTinterwordspacing
S.~Verdoolaege, J.~Carlos~Juega, A.~Cohen, J.~Ignacio~G\'{o}mez, C.~Tenllado,
  and F.~Catthoor, ``Polyhedral parallel code generation for {CUDA},''
  \emph{ACM Trans. Archit. Code Optim.}, vol.~9, no.~4, Jan. 2013. [Online].
  Available: \url{https://doi.org/10.1145/2400682.2400713}
\BIBentrySTDinterwordspacing

\bibitem{zinenko2018scheduling}
\BIBentryALTinterwordspacing
O.~Zinenko, S.~Verdoolaege, C.~Reddy, J.~Shirako, T.~Grosser, V.~Sarkar, and
  A.~Cohen, ``Modeling the conflicting demands of parallelism and
  temporal/spatial locality in affine scheduling,'' in \emph{Proceedings of the
  27th International Conference on Compiler Construction}, ser. CC 2018.\hskip
  1em plus 0.5em minus 0.4em\relax New York, NY, USA: Association for Computing
  Machinery, 2018, pp. 3--13. [Online]. Available:
  \url{https://doi.org/10.1145/3178372.3179507}
\BIBentrySTDinterwordspacing

\bibitem{grosser2012polly}
T.~Grosser, A.~Groesslinger, and C.~Lengauer, ``Polly—performing polyhedral
  optimizations on a low-level intermediate representation,'' \emph{Parallel
  Processing Letters}, vol.~22, no.~04, p. 1250010, 2012.

\bibitem{vasilache2006violated}
N.~Vasilache, C.~Bastoul, A.~Cohen, and S.~Girbal, ``Violated dependence
  analysis,'' in \emph{Proceedings of the 20th annual international conference
  on Supercomputing}, 2006, pp. 335--344.

\bibitem{beaugnon2017optimization}
U.~Beaugnon, A.~Pouille, M.~Pouzet, J.~Pienaar, and A.~Cohen, ``Optimization
  space pruning without regrets,'' in \emph{Proceedings of the 26th
  International Conference on Compiler Construction}, 2017, pp. 34--44.

\bibitem{adams2019halideopt}
\BIBentryALTinterwordspacing
A.~Adams, K.~Ma, L.~Anderson, R.~Baghdadi, T.-M. Li, M.~Gharbi, B.~Steiner,
  S.~Johnson, K.~Fatahalian, F.~Durand, and J.~Ragan-Kelley, ``Learning to
  optimize halide with tree search and random programs,'' \emph{ACM Trans.
  Graph.}, vol.~38, no.~4, Jul. 2019. [Online]. Available:
  \url{https://doi.org/10.1145/3306346.3322967}
\BIBentrySTDinterwordspacing

\bibitem{gysi2019absinthe}
T.~{Gysi}, T.~{Grosser}, and T.~{Hoefler}, ``Absinthe: Learning an analytical
  performance model to fuse and tile stencil codes in one shot,'' in \emph{2019
  28th International Conference on Parallel Architectures and Compilation
  Techniques (PACT)}, 2019, pp. 370--382.

\bibitem{gysi2015modesto}
\BIBentryALTinterwordspacing
T.~Gysi, T.~Grosser, and T.~Hoefler, ``{MODESTO}: Data-centric analytic
  optimization of complex stencil programs on heterogeneous architectures,'' in
  \emph{Proceedings of the 29th ACM on International Conference on
  Supercomputing}, ser. ICS '15.\hskip 1em plus 0.5em minus 0.4em\relax New
  York, NY, USA: Association for Computing Machinery, 2015, pp. 177--186.
  [Online]. Available: \url{https://doi.org/10.1145/2751205.2751223}
\BIBentrySTDinterwordspacing

\bibitem{cosmoorg}
``Consortium for small-scale modeling,'' \url{http://www.cosmo-model.org/},
  2020.

\bibitem{fv3}
``{FV3}: Finite-volume cubed-sphere dynamical core,''
  \url{https://www.gfdl.noaa.gov/fv3/}, 2020.

\bibitem{osuna2020dawn}
\BIBentryALTinterwordspacing
C.~Osuna, T.~Wicky, F.~Thuering, T.~Hoefler, and O.~Fuhrer, ``Dawn: a
  high-level domain-specific language compiler toolchain for weather and
  climate applications,'' \emph{Supercomputing Frontiers and Innovations},
  vol.~7, no.~2, 2020. [Online]. Available:
  \url{https://www.superfri.org/superfri/article/view/314}
\BIBentrySTDinterwordspacing

\bibitem{cuda2008}
\BIBentryALTinterwordspacing
J.~Nickolls, I.~Buck, M.~Garland, and K.~Skadron, ``Scalable parallel
  programming with {CUDA},'' \emph{Queue}, vol.~6, no.~2, pp. 40--53, Mar.
  2008. [Online]. Available: \url{https://doi.org/10.1145/1365490.1365500}
\BIBentrySTDinterwordspacing

\bibitem{holewinski2012overlapped}
\BIBentryALTinterwordspacing
J.~Holewinski, L.-N. Pouchet, and P.~Sadayappan, ``High-performance code
  generation for stencil computations on {GPU} architectures,'' in
  \emph{Proceedings of the 26th ACM International Conference on
  Supercomputing}, ser. ICS '12.\hskip 1em plus 0.5em minus 0.4em\relax New
  York, NY, USA: Association for Computing Machinery, 2012, pp. 311--320.
  [Online]. Available: \url{https://doi.org/10.1145/2304576.2304619}
\BIBentrySTDinterwordspacing

\bibitem{rawat2016streaming}
\BIBentryALTinterwordspacing
P.~S. Rawat, C.~Hong, M.~Ravishankar, V.~Grover, L.-N. Pouchet, and
  P.~Sadayappan, ``Effective resource management for enhancing performance of
  2d and 3d stencils on {GPUs},'' in \emph{Proceedings of the 9th Annual
  Workshop on General Purpose Processing Using Graphics Processing Unit}, ser.
  GPGPU '16.\hskip 1em plus 0.5em minus 0.4em\relax New York, NY, USA:
  Association for Computing Machinery, 2016, pp. 92--102. [Online]. Available:
  \url{https://doi.org/10.1145/2884045.2884047}
\BIBentrySTDinterwordspacing

\bibitem{gt4pygithub}
``{GT4Py},'' \url{https://github.com/gridtools/gt4py}, 2020.

\bibitem{edwards2013kokkos}
H.~C. {Edwards} and C.~R. {Trott}, ``Kokkos: Enabling performance portability
  across manycore architectures,'' in \emph{2013 Extreme Scaling Workshop (xsw
  2013)}, 2013, pp. 18--24.

\bibitem{rajagithub}
``{RAJA},'' \url{https://github.com/LLNL/RAJA}, 2020.

\bibitem{baghdadi2015pencil}
R.~{Baghdadi}, U.~{Beaugnon}, A.~{Cohen}, T.~{Grosser}, M.~{Kruse}, C.~{Reddy},
  S.~{Verdoolaege}, A.~{Betts}, A.~F. {Donaldson}, J.~{Ketema}, J.~{Absar},
  S.~v.~{Haastregt}, A.~{Kravets}, A.~{Lokhmotov}, R.~{David}, and
  E.~{Hajiyev}, ``{PENCIL}: A platform-neutral compute intermediate language
  for accelerator programming,'' in \emph{2015 International Conference on
  Parallel Architecture and Compilation (PACT)}, 2015, pp. 138--149.

\bibitem{bennun2019dace}
T.~Ben-Nun, J.~de~Fine~Licht, A.~N. Ziogas, T.~Schneider, and T.~Hoefler,
  ``Stateful dataflow multigraphs: A data-centric model for performance
  portability on heterogeneous architectures,'' in \emph{Proceedings of the
  International Conference for High Performance Computing, Networking, Storage
  and Analysis}, ser. SC '19, 2019.

\bibitem{hagedorn2018liftstencils}
\BIBentryALTinterwordspacing
B.~Hagedorn, L.~Stoltzfus, M.~Steuwer, S.~Gorlatch, and C.~Dubach, ``High
  performance stencil code generation with {Lift},'' in \emph{Proceedings of
  the 2018 International Symposium on Code Generation and Optimization}, ser.
  CGO 2018.\hskip 1em plus 0.5em minus 0.4em\relax New York, NY, USA:
  Association for Computing Machinery, 2018, pp. 100--112. [Online]. Available:
  \url{https://doi.org/10.1145/3168824}
\BIBentrySTDinterwordspacing

\bibitem{climagithub}
``{CLIMA},'' \url{https://github.com/climate-machine/CLIMA/}, 2020.

\bibitem{adams2019lfric}
\BIBentryALTinterwordspacing
S.~Adams, R.~Ford, M.~Hambley, J.~Hobson, I.~Kav\u{c}i\u{c}, C.~Maynard,
  T.~Melvin, E.~M\"uller, S.~Mullerworth, A.~Porter, M.~Rezny, B.~Shipway, and
  R.~Wong, ``{LFRic}: Meeting the challenges of scalability and performance
  portability in weather and climate models,'' \emph{Journal of Parallel and
  Distributed Computing}, vol. 132, pp. 383 -- 396, 2019. [Online]. Available:
  \url{http://www.sciencedirect.com/science/article/pii/S0743731518305306}
\BIBentrySTDinterwordspacing

\bibitem{gridtoolsgithub}
``{GridTools},'' \url{https://github.com/GridTools/gridtools}, 2020.

\bibitem{clement2018claw}
\BIBentryALTinterwordspacing
V.~Clement, S.~Ferrachat, O.~Fuhrer, X.~Lapillonne, C.~E. Osuna, R.~Pincus,
  J.~Rood, and W.~Sawyer, ``The {CLAW DSL}: Abstractions for performance
  portable weather and climate models,'' in \emph{Proceedings of the Platform
  for Advanced Scientific Computing Conference}, ser. PASC '18.\hskip 1em plus
  0.5em minus 0.4em\relax New York, NY, USA: Association for Computing
  Machinery, 2018. [Online]. Available:
  \url{https://doi.org/10.1145/3218176.3218226}
\BIBentrySTDinterwordspacing

\bibitem{mueller2018hybridfortran}
M.~M{\"u}ller and T.~Aoki, ``{Hybrid Fortran}: High productivity {GPU} porting
  framework applied to japanese weather prediction model,'' in
  \emph{Accelerator Programming Using Directives}, S.~Chandrasekaran and
  G.~Juckeland, Eds.\hskip 1em plus 0.5em minus 0.4em\relax Cham: Springer
  International Publishing, 2018, pp. 20--41.

\bibitem{leissa2018anydsl}
\BIBentryALTinterwordspacing
R.~Lei\ss{}a, K.~Boesche, S.~Hack, A.~P\'{e}rard-Gayot, R.~Membarth,
  P.~Slusallek, A.~M\"{u}ller, and B.~Schmidt, ``{AnyDSL}: A partial evaluation
  framework for programming high-performance libraries,'' \emph{Proc. ACM
  Program. Lang.}, vol.~2, no. OOPSLA, Oct. 2018. [Online]. Available:
  \url{https://doi.org/10.1145/3276489}
\BIBentrySTDinterwordspacing

\bibitem{rompf2010staging}
\BIBentryALTinterwordspacing
T.~Rompf and M.~Odersky, ``Lightweight modular staging: A pragmatic approach to
  runtime code generation and compiled {DSLs},'' in \emph{Proceedings of the
  Ninth International Conference on Generative Programming and Component
  Engineering}, ser. GPCE '10.\hskip 1em plus 0.5em minus 0.4em\relax New York,
  NY, USA: Association for Computing Machinery, 2010, pp. 127--136. [Online].
  Available: \url{https://doi.org/10.1145/1868294.1868314}
\BIBentrySTDinterwordspacing

\bibitem{devito2013terra}
\BIBentryALTinterwordspacing
Z.~DeVito, J.~Hegarty, A.~Aiken, P.~Hanrahan, and J.~Vitek, ``Terra: A
  multi-stage language for high-performance computing,'' in \emph{Proceedings
  of the 34th ACM SIGPLAN Conference on Programming Language Design and
  Implementation}, ser. PLDI '13.\hskip 1em plus 0.5em minus 0.4em\relax New
  York, NY, USA: Association for Computing Machinery, 2013, pp. 105--116.
  [Online]. Available: \url{https://doi.org/10.1145/2491956.2462166}
\BIBentrySTDinterwordspacing

\bibitem{steuwer2017lift}
M.~{Steuwer}, T.~{Remmelg}, and C.~{Dubach}, ``{LIFT}: A functional
  data-parallel ir for high-performance {GPU} code generation,'' in \emph{2017
  IEEE/ACM International Symposium on Code Generation and Optimization (CGO)},
  2017, pp. 74--85.

\bibitem{nguyen2010blocking}
A.~{Nguyen}, N.~{Satish}, J.~{Chhugani}, C.~{Kim}, and P.~{Dubey}, ``{3.5-D}
  blocking optimization for stencil computations on modern {CPUs} and {GPUs},''
  in \emph{SC '10: Proceedings of the 2010 ACM/IEEE International Conference
  for High Performance Computing, Networking, Storage and Analysis}, 2010, pp.
  1--13.

\bibitem{maruyama2014optimizing}
N.~Maruyama and T.~Aoki, ``Optimizing stencil computations for {NVIDIA} kepler
  {GPUs},'' in \emph{Proceedings of the 1st international workshop on
  high-performance stencil computations, Vienna}, 2014, pp. 89--95.

\bibitem{grosser2014hybrid}
\BIBentryALTinterwordspacing
T.~Grosser, A.~Cohen, J.~Holewinski, P.~Sadayappan, and S.~Verdoolaege,
  ``Hybrid hexagonal/classical tiling for {GPUs},'' in \emph{Proceedings of
  Annual IEEE/ACM International Symposium on Code Generation and Optimization},
  ser. CGO '14.\hskip 1em plus 0.5em minus 0.4em\relax New York, NY, USA:
  Association for Computing Machinery, 2014, pp. 66--75. [Online]. Available:
  \url{https://doi.org/10.1145/2544137.2544160}
\BIBentrySTDinterwordspacing

\bibitem{grosser2014split}
\BIBentryALTinterwordspacing
T.~Grosser, A.~Cohen, P.~H.~J. Kelly, J.~Ramanujam, P.~Sadayappan, and
  S.~Verdoolaege, ``Split tiling for {GPUs}: Automatic parallelization using
  trapezoidal tiles,'' in \emph{Proceedings of the 6th Workshop on General
  Purpose Processor Using Graphics Processing Units}, ser. GPGPU-6.\hskip 1em
  plus 0.5em minus 0.4em\relax New York, NY, USA: Association for Computing
  Machinery, 2013, pp. 24--31. [Online]. Available:
  \url{https://doi.org/10.1145/2458523.2458526}
\BIBentrySTDinterwordspacing

\bibitem{matsumura2020an5d}
\BIBentryALTinterwordspacing
K.~Matsumura, H.~R. Zohouri, M.~Wahib, T.~Endo, and S.~Matsuoka, ``{AN5D}:
  Automated stencil framework for high-degree temporal blocking on {GPUs},'' in
  \emph{Proceedings of the 18th ACM/IEEE International Symposium on Code
  Generation and Optimization}, ser. CGO 2020.\hskip 1em plus 0.5em minus
  0.4em\relax New York, NY, USA: Association for Computing Machinery, 2020, pp.
  199--211. [Online]. Available: \url{https://doi.org/10.1145/3368826.3377904}
\BIBentrySTDinterwordspacing

\bibitem{rawat2018fusion}
P.~S. {Rawat}, M.~{Vaidya}, A.~{Sukumaran-Rajam}, M.~{Ravishankar},
  V.~{Grover}, A.~{Rountev}, L.~{Pouchet}, and P.~{Sadayappan},
  ``Domain-specific optimization and generation of high-performance {GPU} code
  for stencil computations,'' \emph{Proceedings of the IEEE}, vol. 106, no.~11,
  pp. 1902--1920, 2018.

\bibitem{wahib2014fusion}
M.~{Wahib} and N.~{Maruyama}, ``Scalable kernel fusion for memory-bound {GPU}
  applications,'' in \emph{SC '14: Proceedings of the International Conference
  for High Performance Computing, Networking, Storage and Analysis}, 2014, pp.
  191--202.

\bibitem{rawat2018regopt}
\BIBentryALTinterwordspacing
P.~S. Rawat, F.~Rastello, A.~Sukumaran-Rajam, L.-N. Pouchet, A.~Rountev, and
  P.~Sadayappan, ``Register optimizations for stencils on {GPUs},'' in
  \emph{Proceedings of the 23rd ACM SIGPLAN Symposium on Principles and
  Practice of Parallel Programming}, ser. PPoPP '18.\hskip 1em plus 0.5em minus
  0.4em\relax New York, NY, USA: Association for Computing Machinery, 2018, pp.
  168--182. [Online]. Available: \url{https://doi.org/10.1145/3178487.3178500}
\BIBentrySTDinterwordspacing

\bibitem{zhao2019stencil}
\BIBentryALTinterwordspacing
T.~Zhao, P.~Basu, S.~Williams, M.~Hall, and H.~Johansen, ``Exploiting reuse and
  vectorization in blocked stencil computations on {CPUs} and {GPUs},'' in
  \emph{Proceedings of the International Conference for High Performance
  Computing, Networking, Storage and Analysis}, ser. SC '19.\hskip 1em plus
  0.5em minus 0.4em\relax New York, NY, USA: Association for Computing
  Machinery, 2019. [Online]. Available:
  \url{https://doi.org/10.1145/3295500.3356210}
\BIBentrySTDinterwordspacing

\end{thebibliography}

\end{document}